\documentclass[12pt,preprint]{aastex}

\usepackage{amsmath}

\slugcomment{To appear in PASP}

\shorttitle{Photonic Passbands and Zero-points}
\shortauthors{Bessell \& Murphy}

\begin{document}


\title{Spectrophotometric Libraries, Revised Photonic Passbands and Zero-points for UBVRI, Hipparcos and Tycho Photometry. }


\author{Michael Bessell \& Simon Murphy }
\affil{RSAA, Mount Stromlo Observatory, The Australian National University, ACT 2611, Australia}
\email{bessell@mso.anu.edu.au}

\begin{abstract}
We have calculated improved photonic passbands for the $UBVRI$, Hipparcos and Tycho $H$$_{p}$, $B$$_{T}$, $V$$_{T}$ standard systems using the extensive spectrophotometric libraries of NGSL and MILES. Using the $H_{p}$ passband, we adjusted the absolute flux levels of stars in the spectrophotometric libraries so their synthetic H$_{p}$ magnitudes matched the precise Hipparcos catalog value. Synthetic photometry based on the renormalized fluxes were compared to the standard $UBVRI$ and $B_{T}$, $V_{T}$ magnitudes and revised synthetic zero-points  were determined. The Hipparcos and Tycho photometry system zero-points were also compared to the $V$ magnitude zero-points of the SAAO $UBVRI$ system, the homogenized $UBV$ system and the Walraven $VB$ system.  The confusion in the literature concerning broadband magnitudes, fluxes, passbands and the choice of appropriate mean wavelengths is detailed and discussed in an appendix.  
\end{abstract}


\newcommand{\bigint}{\displaystyle \int }
\section{Introduction}
The Hipparcos Catalogue \citep{Perr97} is a high precision photometric (plus parallax and proper-motion) catalog of more than 100,000 stars measured with the $H_{p}$ band; the Tycho2 catalog \citep{Hog00} contains 2.5 million stars measured (mostly) with lower precision in the $B_{T}$ and $V_{T}$ bands. The remarkable collection of data was obtained during the 4 year (1989--1993) mission of the Hipparcos satellite. The Hipparcos and Tycho photometric systems and their measured median precisions were discussed by \citet{vLee97b} and the passbands were given in \citet{vLee97a}. However, the detectors suffered degradation throughout the mission as a result of being launched into an incorrect orbit and this degradation invalidated the measured pre-launch passbands.  \citet{Bess00} devised self-consistent Hipparcos and Tycho passbands by comparing regressions of $V-H_{p}$, $V-B_{T}$ and $V-V_{T}$ versus  $V-I$ for a sample of precise E-region $UBVRI$ standard stars with synthetic photometry computed from the $R\sim$100 Vilnius averaged spectra \citep{Stra72}. This indicated the necessity of a significant redward shift of the blue edge of the published $H_{p}$ band but only small changes for the Tycho bands. However, the passbands may not have been definitive because of the small number of averaged Vilnius spectra used and their low resolution. 

In the last few years, two libraries of accurate higher resolution ($R\sim$1000--2000) spectrophotometric data have become available -  the Next Generation Spectral Library (NGSL) \citep[][\url{http://archive.stsci.edu/prepds/stisngsl/index.html}]{Heap07} and the Medium Resolution INT Library of Empirical Spectra (MILES) \citep[][\url{http://www.iac.es/proyecto/miles/}]{Sanc06}. Many of the stars in these spectral libraries also have Hipparcos and Tycho magnitudes - providing the opportunity to reexamine the passbands of the Hipparcos and Tycho systems.  Furthermore, the high precision of the Hipparcos magnitudes \citep[see section 1.3.1][]{Perr97} enables them to be used to adjust the flux levels of the data in the NGSL libraries and make the stars extremely valuable for whole sky spectrophotometric calibration of imaging surveys such as SkyMapper \citep{Kell07}.     

Being space based, a unique property of the Hipparcos photometric systems is the absence of any seasonal or hemisphere-related effects seen in some ground based photometric systems due to variations in temperature, atmospheric extinction and instrumental orientation. The Hipparcos photometry database can therefore be compared with databases of ground-based photometric systems to examine their magnitude zero-points and to look for any systematic offsets in the photometry, as discussed by \citet{vLee97a} and \citet{Pell07}.

In this paper, we will outline the derivation of improved $UBVRI$, Hipparcos and Tycho passbands by using synthetic photometry from spectrophotometric atlases and comparing it with broad-band photometry. We will also adjust the absolute levels of the spectrophotometric fluxes by comparing the synthesized $H_{p}$ magnitudes with the Hipparcos catalogue magnitudes. In addition, we will use the mean differences between the synthetic and the observed photometry to determine zero-point corrections for the $UBVRI$, $B$$_{T}$ and $V$$_{T}$ bands.  We will also inter-compare the zero-points of the SAAO $UBVRI$, the homogenized $UBV$ and the Walraven $VB$ systems. Finally, in an appendix we discuss confusion and inexactness concerning the derivation of mean fluxes, response functions and the plethora of expressions for mean wavelengths and frequencies associated with broadband photometry.  

\section{Synthetic Photometry}
The synthetic photometry in this paper was computed using two photometry packages, one written by Andrew Pickles in 1980, the other \textit{pysynphot}\footnotemark  (\url{http://stsdas.stsci.edu/pysynphot}). The synthetic photometry was computed by evaluating, for each band '$x$', the expression
\footnotetext{Version 0.9 distributed as part of stsci-python 2.12 (Aug 2011)}
\begin{equation} 
$$ mag$_{x}$ = AB $-$ ZP$_{x}$  $$
\end{equation}
where
\begin{equation}
$$ AB =   $-2.5 \log\frac{\bigint{f_{\nu}(\nu) S_{x}(\nu)  d\nu/\nu}} {\bigint{S_{x}(\nu) d\nu/\nu}} - 48.60$ =  $-2.5 \log\frac{\bigint{f_{\lambda}(\lambda) S_{x}(\lambda) \lambda d\lambda}} { \bigint{S_{x}(\lambda)c\,d\lambda /{\lambda}}} - 48.60$     $$
\end{equation}
and $f_{\nu}(\nu)$ is the observed absolute flux in erg cm$^{-2}$ sec$^{-1}$ Hz$^{-1}$, $f_{\lambda}(\lambda)$ is the observed absolute flux in erg cm$^{-2}$ sec$^{-1}$ \AA$^{-1}$,  $S_{x}(\lambda)$ are the photonic passbands  (response functions), $\lambda$ is the wavelength in \AA, and ZP$_{x}$ are the zero-point magnitudes for each band   (see Sections 5.4, 7 and Appendix). For SI units the constants in the above equations would be different as an erg cm$^{-2}$ sec$^{-1}$ is equivalent to 10$^{-3}$ W m$^{-2}$. 

For accurate photometry it is important that the passbands provided to the integration routines are well sampled and smooth. Because passbands are usually published at coarse wavelength intervals (25--100\AA), it is necessary to interpolate these passband tables to a finer spacing of a few \AA\ using a univariate spline or a parabolic interpolation routine.  The physical passbands themselves are smooth and the recommended interpolation recovers this. Our two packages produced identical results after this step.  

\section{Complications and caveats to the realization of standard systems}
There is a fundamental concern associated with the theoretical realization of 
the older evolved standard photometric systems in order to produce synthetic 
photometry from theoretical and obervational fluxes. The technique used 
is to reverse engineer the standard system's passband sensitivity 
functions by comparing synthetic photometry with observations \citep[e.g.][]{Strai96},\citep[$UBVRIJHKL$:][] {Bes90a, Bess88}. That is, commencing with a 
passband based on an author's prescription of detector and filter bandpass, 
synthetic magnitudes are computed from absolute or relative absolute 
spectrophotometric fluxes for stars with known standard colors. By slightly 
modifying the initial passband (shifting the central wavelength or altering 
the blue or red cutoff) and recomputing the synthetic colors, it is usually 
possible to devise a bandpass that generates magnitudes that differ from the 
standard magnitudes \emph {within the errors} by only a constant that is 
independent of the color of the star. It is usually taken for granted that 
such a unique passband exists and that given a large enough set of precise 
spectrophotometric data and sufficient passband adjustment trials, it can be 
recovered. However, there are several reasons why this may not be the case, 
at least not across the complete temperature range.

Whilst the original system may have been based on a real set of filters and
detector, the original set of standard stars would almost certainly have
been obtained with lower precision than is now possible and for stars of a
restricted temperature and luminosity range. The filters may also have been
replaced during the establishment of the system and the later data linearly
transformed onto mean relations shown by the previous data. In addition, the
contemporary lists of very high precision secondary standards that
essentially define the ``standard systems" have all been measured using
more sensitive equipment, with different wavelength responses. Again, rather
than preserve the natural scale of the contemporary equipment the
measurements have been ``transformed" to some mean representation of the
original system by applying one or more linear transformations or even
non-linear transformations \citep[e.g.][]{Menz93}. To incorporate bluer or redder 
stars than those in the original standard lists \citep[e.g.][]{Kilk98}, 
extrapolations have also been made and these
may have been unavoidably skewed by the imprecision of the original data and
the small number of stars with extreme colors in the original lists. As a
result, the contemporary standard system, \emph {although well defined
observationally} by lists of stars with precise colors and magnitudes, \emph {may
not represent any real system} and is therefore impossible to realize
with a unique passband that can reproduce the standard magnitudes and colors
through a linear transformation with a slope of 1.0. 

In fact, perhaps we should not be trying to find a unique passband with a 
central wavelength and shape that can reproduce the colors of a standard 
system but rather we should be trying to match the passbands and the linear 
(but non-unity slope) or non-linear transformations used by the contemporary 
standard system authors to transform their natural photometry onto the 
``standard system". The revised realization of the Geneva photometric system by \citet{Nico96} uses this philosophy, as has one of us (MSB) in a forthcoming paper on the realization of the $uvby$ system. 

However, in this paper we have set out in the traditional way, as outlined above, to adjust the passbands to achieve agreement between the synthetic photometry and the standard system photometry within the errors of the standard system. It may seem desirable to do these passband adjustments in a less ad hoc way, but given the uncertainties underlying existing standard system photometry a more accurate method is unnecessary, at present.   

\section{The NGSL and MILES spectra}
Because of how spectroscopic observations are normally made, spectrophotometric fluxes are calibrated mainly to determine the accurate relative absolute fluxes (the flux variation with wavelength) but not the absolute flux (the apparent magnitude). Depending on the slit width and the seeing or other instrumental effects, the resultant absolute flux levels may be measured only to a precision of 0.1--0.2 mags. To assign an accurate absolute flux level one is normally required to compute a synthetic magnitude from the spectrophotometric fluxes and equate this to a standard magnitude for the object, often an existing  magnitude, such as $B$ or $V$. Currently, the most precise magnitudes available for the largest number of stars are the Hipparcos $H_{p}$ magnitudes. There are 72,300 stars in the Hipparcos Catalogue with median $H_{p}$ magnitudes given to better than 0.002 mag. We have therefore synthesized the $H_{p}$ magnitudes for all the stars in the spectrophotometric libraries and adjusted the absolute flux scale of each star (that is in the Hipparcos catalogue) to match the catalogue $H_{p}$ value, thus producing spectrophotometric data with an uncertainty in the overall absolute flux level of a few millimags. In Fig 1 we show histograms of the differences for the NGSL stars between the observed and synthetic $V$ mags before and after renormalization to the $H_{p}$ mags. We could not show this comparison for the MILES spectra as they published only relative-absolute fluxes (normalized to 1 at  4500\AA), but after renormalization to the $H_{p}$ magnitudes, the delta $V$ distributions for both NGSL and MILES spectral libraries are approximately gaussian with a similar rms of 0.017 mags. 

\begin{figure}
\figurenum{1}
\epsscale{.80}
\plotone{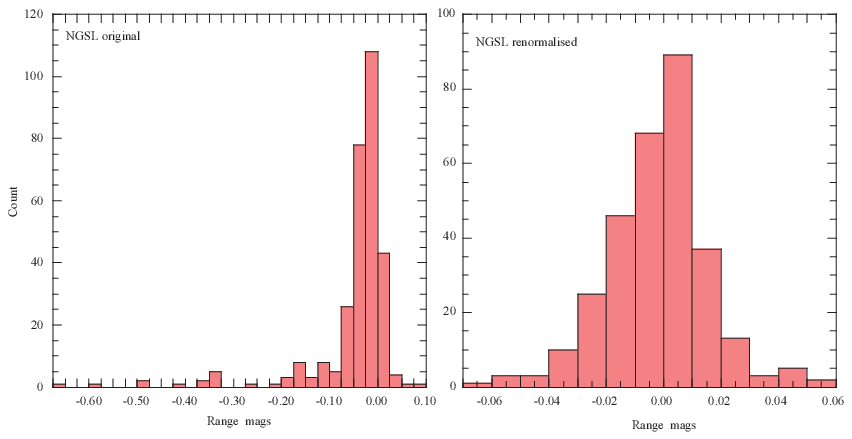}
\caption{Histogram of difference between observed and synthetic $V$ magnitudes for original NGSL spectra (left)  and renormalized spectra (right). The renormalized MILES spectra show a very similar distribution and rms.} 
\end{figure}

The wavelength range of the NGSL spectra encompasses the wavelength range of  the Hipparcos and Tycho passbands, but because the MILES spectra do not cover the complete extent of the red tails of the $H_{p}$ and the $R$ band, nor any of the $I$ band, we have extrapolated the MILES spectra from 7000\AA\  to 9900\AA\ using model atmosphere fits to the 3500\AA\--7000\AA\ region by \citet{Kerz11}. The grids used were from ATLAS \citep{Muna05} for $T_{eff}>$ 8000K and MARCS \citep{Gust08} for 8000K$>T_{eff}>$ 2500K. The MILES spectra were also extrapolated from 3500\AA\ to 3000\AA\ to cover the $U$ band.  These extrapolations  may result in a slight uncertainty in the synthetic photometry in some passbands from the MILES spectra; however, we think that it is small, as shown by the insignificant differences between the relations using the NGSL and MILES spectra, except probably for the M stars.  

The 373 adjusted NGSL spectrophotometric fluxes, covering the wavelength range 1800\AA\ to 10100\AA\, and with a precise absolute flux level, are ideally suited to calibrate whole sky surveys, such as SkyMapper \citep{Kell07} and PanSTARRS \citep{Kais10}. These revised absolute fluxes are available from the authors together with the absolute fluxes for the 836 MILES spectra that have Hipparcos photometry.  

\section{The $UBVRI$ passbands}
The Johnson-Cousins $UBVRI$ system passbands have been well discussed \citep[e.g.][]{Azus69, Busk78, Bes90a}, most recently by  \citet{Maiz06} who reconsidered the $UBV$ passbands. Although accepting the \citet{Bes90a} $BV$ passbands, Maiz Appelaniz suggested an unusual and unphysical $U$ passband as providing a better fit to standard $U$ photometry. However, these previous analyses did not have available the large number of revised spectra in the NGSL and the MILES catalogs.  It is very worthwhile, therefore, to reexamine the $UBVRI$ passbands using synthetic $UBVRI$ photometry derived from these extensive data sets.  Standard system $UBV$ data for most of the MILES and NGSL stars are available in the  homogenized $UBV$ catalog  of \citep{Merm06}. $VI$ data are available for many stars in the Hipparcos catalog, while various data sets of Cousins, Menzies, Landolt, Bessell, Kilkenny and Koen also provided much supplementary $VRI$ data \citep{Cous74, Cous76, Cous84, Cous93, Land83, Land09, Bes90b, Kilk98, Koen02, Koen10} .
 
\begin{figure}
\figurenum{2}
\epsscale{.80}
\plotone{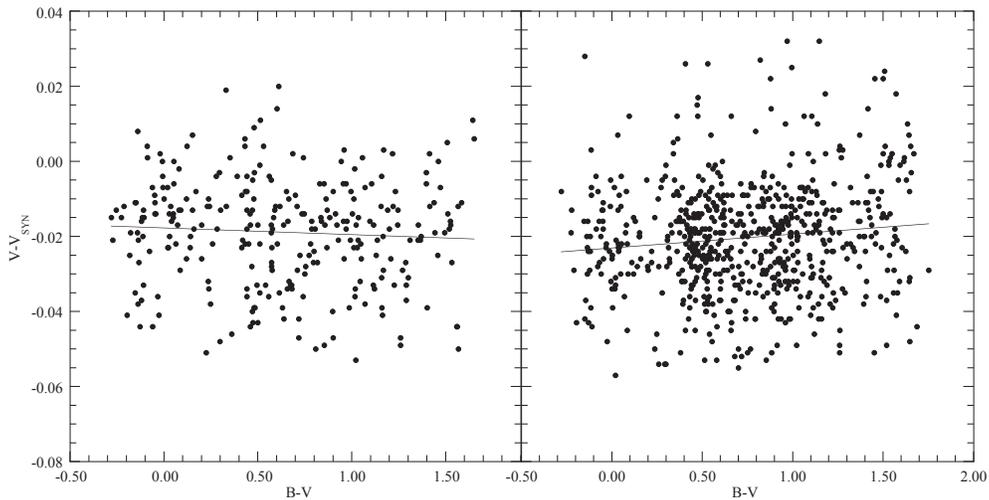}
\caption{Difference between observed and synthetic $V$ magnitudes for NGSL (left)  and MILES (right) samples.   The solid lines show linear fits to the data.} 
\end{figure}

\subsection{The $B$ and $V$ passbands}
After comparing the observed and synthetic photometry, very small slopes were evident in the $\Delta$ $B$ and $\Delta$ $V$ regressions against $B-V$ using the \citet{Bes90a} passbands with the NGSL and MILES samples. These  slopes were removed by making  a small redward shift to the red side of the $V90$ band and a very small overall redward shift in the $B90$ band. The regressions for the MILES sample were similar but not identical. In Fig 2 and Fig 3 we show for the NGSL spectra the differences between the observed and synthetic $V$ and $B-V$, respectively, for our adopted passbands. There are no significant color terms evident, but the synthetic $V$ and $B-V$ mag scales have small apparent offsets associated with the initial adopted zero-points (hereinafter ZPs) of the synthetic photometry. These will be addressed further in Section 7.

\begin{figure}
\figurenum{3}
\epsscale{.80}
\plotone{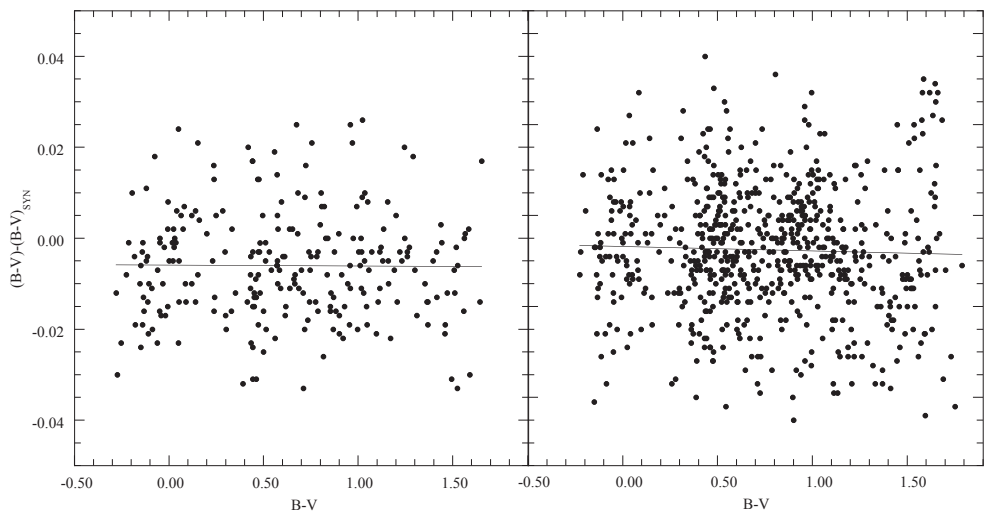}
\caption{Difference between observed and synthetic $B-V$ for NGSL (left)  and MILES (right) samples.   The solid lines show linear fits to the data.} 
\end{figure}

\subsection{The $U$ bandpass revisited}
Standard $U$ photometry has a bad reputation due to the much larger systematic differences in $U-B$ between observers than is evident for $V$ and $B-V$. These systematic differences arise because in stars, $U$ measures the flux across the region of the Balmer Jump and its response is therefore much more sensitive to the exact placement of the band compared to the placement of $B$ and $V$. Many observers take insufficient care to match the position and width of the standard Cousins or Johnson $U$ passband and attempts to standardize the resulting $U-B$ color using a single $B-V$ or $U-B$ color correction term have introduced systematic errors, especially for reddened stars.
\citet{Cous84}  \citep[reprised in][]{Bes90a} outlined such systematic differences evident in different versions of the $U-B$ system.

\begin{figure}
\epsscale{.80}
\figurenum{4}
\plotone{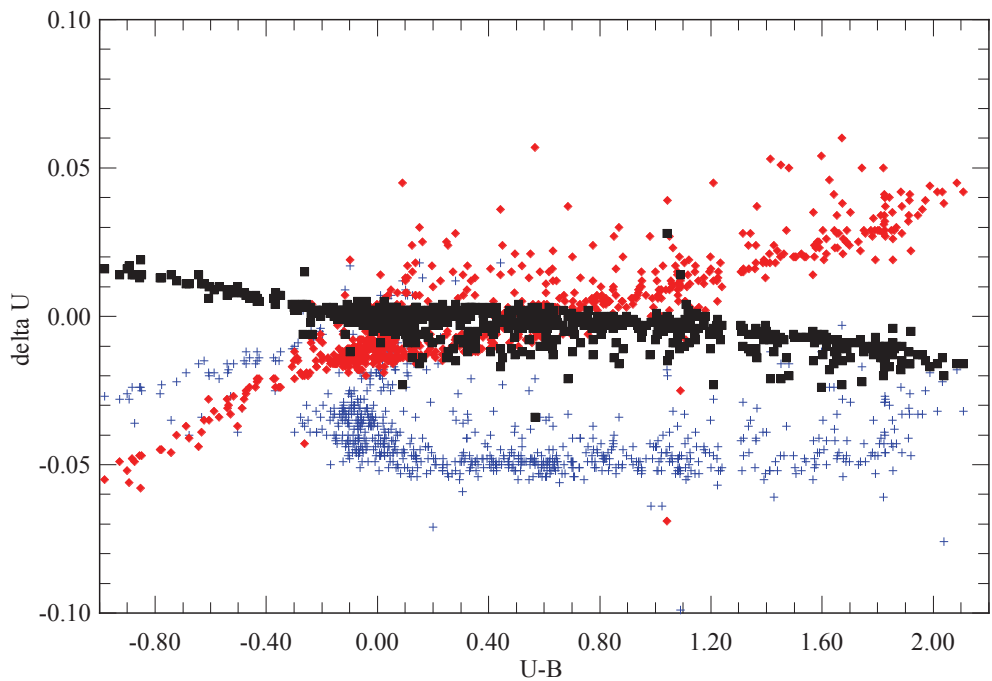}
\caption{Differences in the MILES sample synthetic $U$mags  computed with the $U$ passband in this paper and those of  \citet{Bes90a}:$UX90$ (black squares), \citet{Busk78}:U3 (red diamonds), \citet{Maiz06} (blue crosses). } 
\end{figure}

\citet{Bess86,Bes90a} discussed in detail the likely response function of the $U$ band from first principles and proposed the $UX90$ band as representing  the original band.  \citet[Appendix E3.1]{Bess98} note that the $U-B$ based on this band should be scaled by 0.96.  Although scaling of this order is common in transforming observational systems  \citep[e.g.][]{Menz93,Land83}, there is a notable reluctance to use such terms in computing synthetic photometry.  In spite of the evidence that most standard systems have evolved with nonlinear or bi-linear correction terms \citep[Appendix E1]{Bess98}, most astronomers believe that a passband can be found that reproduces the standard system without the need for linear and/or nonlinear correction terms. In the spirit of that quixotic endeavour, \citet{Busk78} and \citet{Maiz06} proposed $U$ passbands that have almost identical red cutoffs to the $UX90$ band, but different UV cutoffs, thus shifting the effective wavelength of $U$ slightly redward. We have also produced a slightly different $U$ band by moving the UV cutoff of the $UX90$ band slightly redward. This produces an acceptable compromise for the $U$ band that fits the observations reasonably well, although a non-linear fit would be better.  
 
In Fig 4, we show regressions against $U-B$ of the differences between the $U$ photometry synthesized with the passband from this paper and those of \citet{Bes90a}:$UX90$, \citet{Busk78}:U3 and \citet{Maiz06}. The main difference between \citet{Bes90a} and \citet{Busk78} is a small difference in slope, whereas the  \citet{Maiz06} passband mainly produces an offset of about 0.05 mag for G, K and M stars compared to the A and B stars. 

\begin{figure}
\epsscale{.80}
\figurenum{5}
\plotone{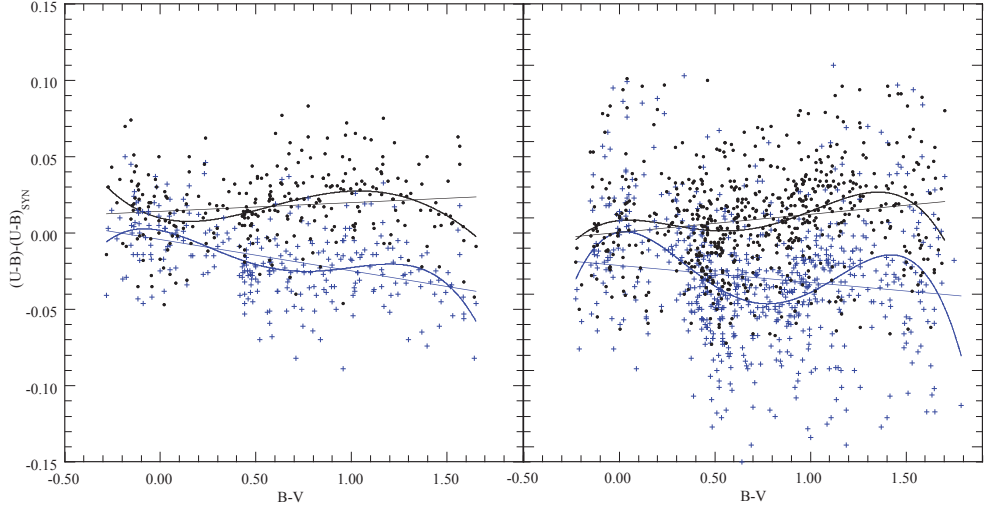}
\caption{Differences between observed $U-B$ and synthetic $U-B$ for the NGSL (left) and MILES (right) sample of stars computed with the $U$ passband in this paper (black dots) and \citet{Maiz06} (blue crosses). The solid lines show the linear and 4th order fits to the differences. } 
\end{figure}

In Fig 5, we show the differences between the observed values of $U-B$ and the synthetic $U-B$ computed for the NGSL and MILES sample of stars. Although the scatter is quite high, the  \citet{Maiz06} $U$ passband clearly does less well and results in a systematic deviation from the standard system for the cooler stars (as anticipated in Fig 4).  

\begin{figure}
\figurenum{6}
\plotone{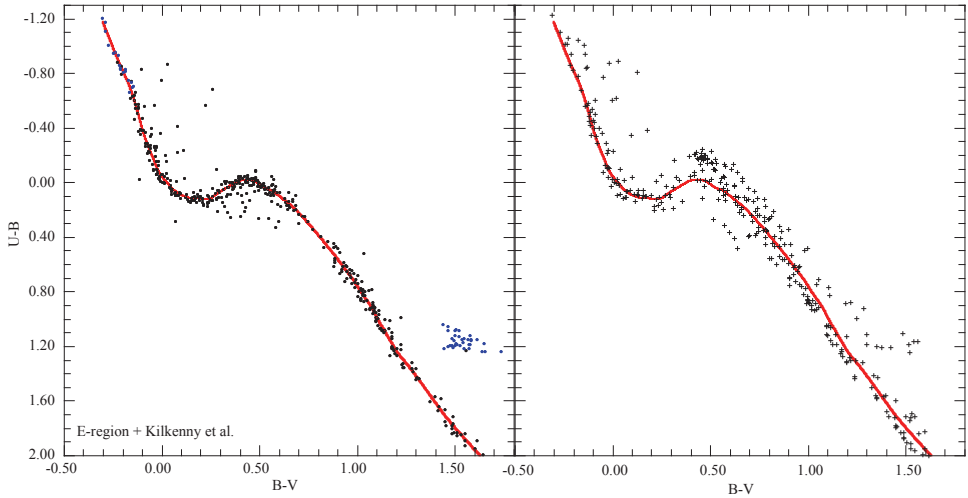}
\caption{Left panel: standard $U-B$ versus $B-V$ relation; E-region stars (black dots), additional blue and red dwarfs (blue dots). Right panel: synthetic $U-B$ versus $B-V$ relation for adopted passbands and the NGSL stars. The thick red line is the same in both figures. The metal deficient stars in the NGSL sample lie above the red fiducial line for $B-V$ between 0.3 and 1.0}
\end{figure}

The SAAO $UBVRI$ photometry \citep{Cous74,Cous76, Kilk98, Koen02, Koen10} represents some of the best standard $UBVRI$ photometry and we use the $U-B$ versus $B-V$ relation from these data (Fig 6 left panel) as the benchmark for comparison with the synthetic photometry. The \citet{Kilk98} photometry (blue dots) extended the standard system to much bluer and redder dwarf stars than represented in the E-region stars (black dots). The red line is a fitted mean line through the O-B-A-F-G dwarf main sequence and the K \& M giants.  In the right panel of Fig 6, the same line is drawn for comparison on the synthetic $U-B$ versus $B-V$ diagram computed for the NGSL sample of stars using our adopted $UBV$ passbands.  Considering that there are many metal-deficient F, G and K stars in the NGSL sample that are not in the empirical sample, the synthetic diagram is in good agreement with the empirical diagram. Note also that most of the reddest stars in the NGSL sample are K and M giants and there are only a few K and M dwarfs.

\subsection{The $R$ and $I$ passbands} 
It was not as straightforward to check the $R$ and $I$ bands because of the lack of precise $RI$ photometry for many of the NGSL and MILES stars. The $V-I$ colors given in the Hipparcos catalog are of uncertain heritage, as are similar data from SIMBAD. An homogenized 
$VRI$ catalog would have been very useful. Our observational data comprised the Hipparcos $V-I$ color supplemented with $VRI$ data mostly from \citet{Bes90b} and \citet{Koen10} for the K and M dwarfs. Although the scatter was high for the Hipparcos $V-I$ comparison, it did indicate that a small shift in the \citet{Bes90a} $I90$ band was needed. We eventually shifted the red edge of the $R90$ band a little redward and the whole $I90$ band a little blueward. In addition to the synthetic photometry from the NGSL and MILES samples, we also had available a small sample of unpublished single observations of mostly late-M dwarfs taken with the DBS at Siding Spring Observatory. As shown in Fig 7 , the resultant synthetic $V-R$ versus $V-I$ relations were in excellent agreement with the empirical loci defined by the precise values in \citet{Menz89}, \citet{Menz90}, \citet{Land09}, \citet{Bes90b} and \citet{Koen10}.  The four empirical data sets are essentially coincident. The sparse redder locus in the $V-R$ versus $V-I$ diagram beyond $V-I\sim 1.8$ are Landolt K and M giants. Note that this and later figures are best viewed magnified in the electronic version.

\begin{figure}
\figurenum{7}
\plotone{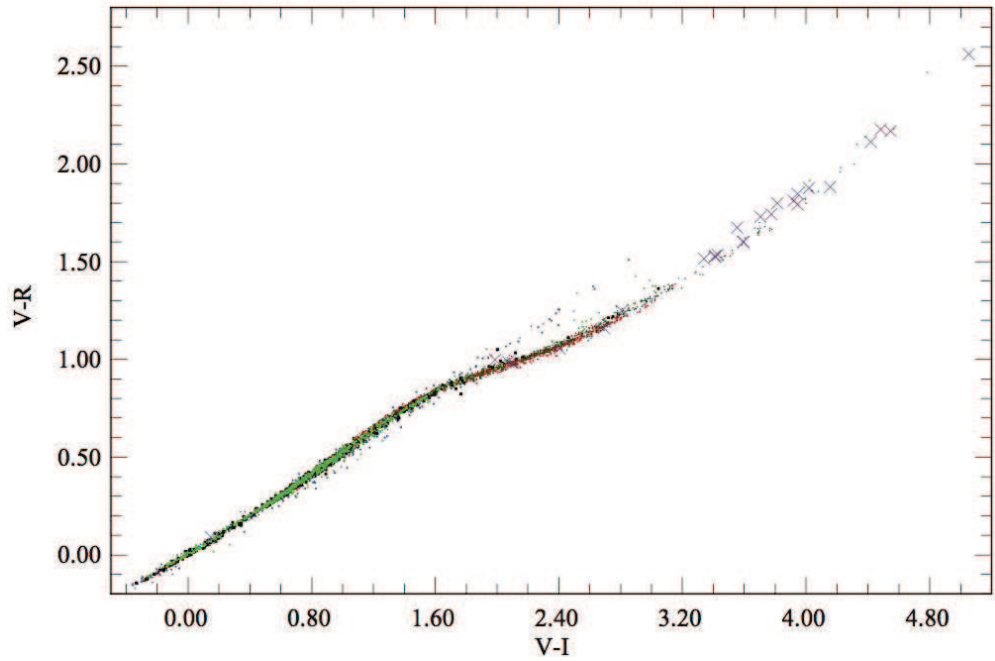}
\caption{Comparison between empirical and synthetic $V-R$ versus $V-I$ relations. Empirical data: \citet{Land09} - blue dots; \citet{Koen10} - red dots; \citet{Bes90b} - dark green pluses. Synthetic data: NGSL - black squares; MILES - light green dots; Bessell 2007 M dwarf spectrophotometry - violet crosses.}
\end{figure}

To better appreciate the comparison we fitted a 9th order polynomial to the empirical $V-R$ versus $V-I$ locus and plotted the $V-R$ residuals of the fit against $V-I$. Applying the same polynomial, we also computed $V-R$ residuals for the synthetic photometry. In Fig 8 we overlay the synthetic residuals, which are seen to agree very well with the trends in the empirical residuals. The few hundredths of a mag systematic differences between the MILES $VRI$ colors for the M dwarfs compared to the empirical stars is undoubtedly due to the extrapolation of the MILES spectra from 7000\AA\ to 9900\AA\ using model spectra. However, for non-M stars, the relation defined by the extrapolated MILES spectra is indistinguishable from the others, indicating an impressive fidelity of the ATLAS \citep{Muna05} and MARCS \citep{Gust08} spectra. 

\begin{figure}
\figurenum{8}
\plotone{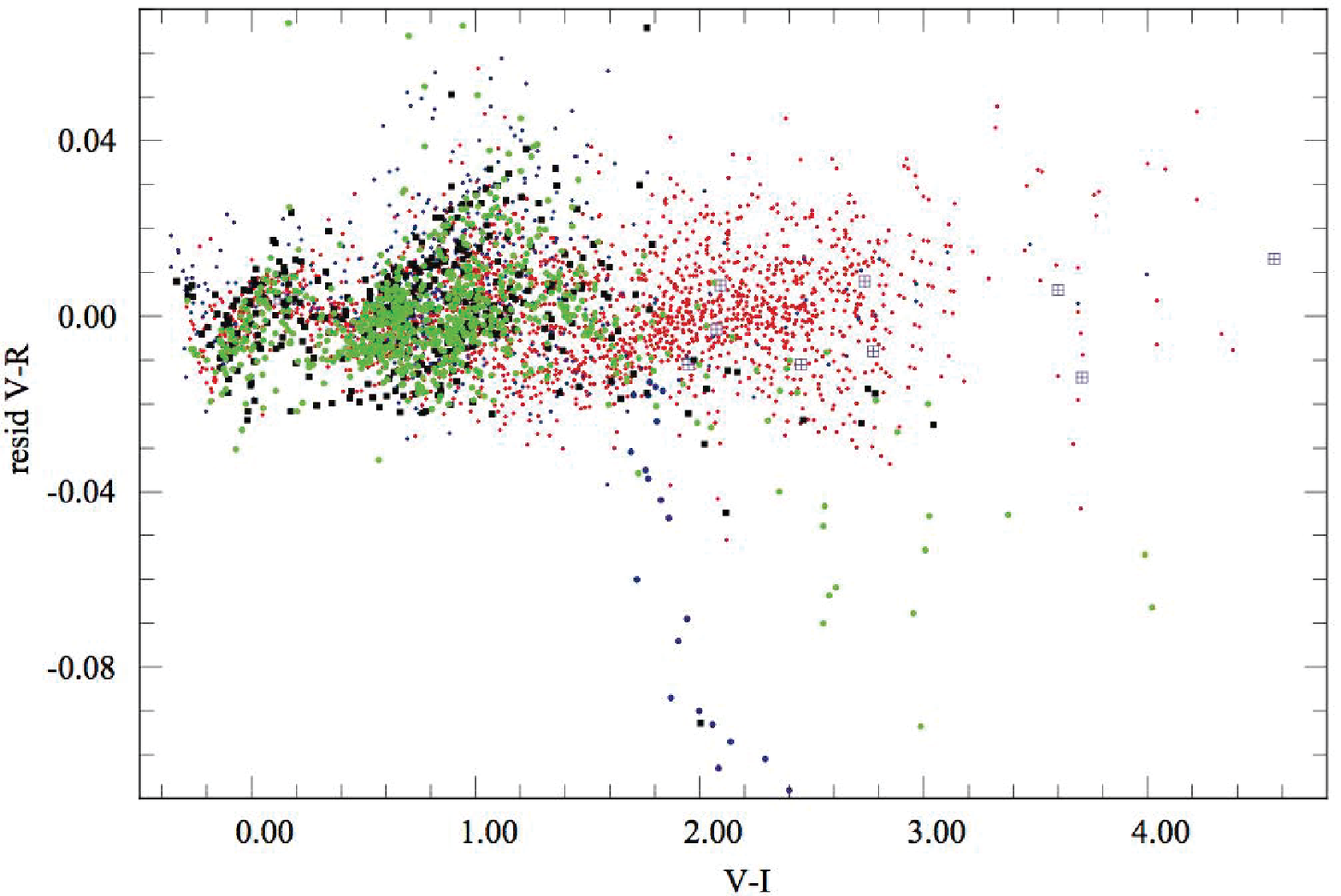}
\caption{Comparison between the residuals of the same 9th order polynomial fit to the synthetic and catalog $V-R$ versus $V-I$ relations (see text for details). Synthetic photometry: NGSL - black squares; MILES - green dots; Bessell unpublished M dwarfs spectra - violet crossed boxes. Observed photometry: E-region stars \citep{Menz89,Menz90} - red dots;  K and M dwarfs \citep{Koen10, Bes90a} - red dots; Landolt \citep{Land09} dwarfs - blue dots, giants - larger blue dots.}
\end{figure}

\subsection{Photometric passbands: photon counting and energy integrating response functions}
There continues to be some confusion in the definition of photometric response functions and their use in computing synthetic photometry. As discussed in \citet{Koor86}, Appendix 4 of \citet{Bess98} and in Appendix A of \citet{Maiz06}, in the era before CCDs,  photometry was largely done with energy measuring detectors. The normalized response functions, $S'_{x} (\lambda)$, that were generally published described the relative fraction of energy detected at different wavelengths across a particular passband. Nowadays, detectors are almost all photon integrating devices, such as CCDs, and the response functions used, $S_{x} (\lambda)$, relate to the relative number of photons detected (or the probability of a photon being detected) at different wavelengths across the passband.  These issues are outlined and explored in the Appendix, where it is also shown why the magnitudes derived from photon counting or energy integration observations are identical (as expected).
In Table 1 we list our adopted normalized photon-counting passbands $S_{x} (\lambda)$ for $U$, $B$, $V$, $R$ and $I$.  In Fig 9 we  show the normalized photon-counting passbands for $U$, $B$, $V$, $R$ and $I$ compared to the \citet{Bes90a} passbands converted to photon-counting. The \citet{Maiz06} $U$ band and the converted \citet{Busk78} photon-counting $U3$ band are also shown.

\begin{figure}
\figurenum{9}
\plotone{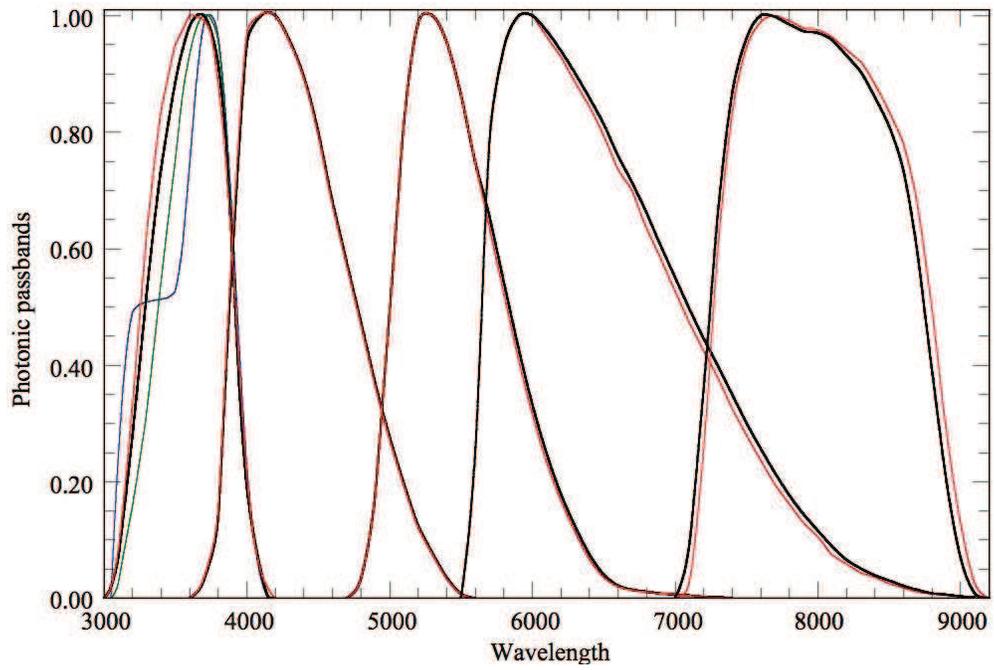}
\caption{Photonic passbands (system response functions)  $S_{x} (\lambda)$ for $UBVRI$. This paper - thick black;   $U$: \citet{Maiz06} - blue;  $U3$: \citet{Busk78} - green;  \citet{Bes90a} - red. See text for details.}
\end{figure}

\section{Hipparcos $H_{p}$ and Tycho $B_{T}$ and $V_{T}$ passbands}
 There are two ground-based photometric systems notable for their precision and stability. These are the Walraven $VBLUW$ photometry of \citet{Pell07} and the $UBVRI$ SAAO and \citet{Land09} photometry discussed above. \citet{Pell90} also provided precise transformations between the Johnson-Cousins $V$ and $B-V$ and Walraven $V$ and $V-B$. 
 
We regressed $V-H_{p}$, $B-B_{T}$ and $V-V_{T}$ versus $B-V$ for these two data sets and compared them with the synthetic photometry from the NGSL and MILES. As done for $UBV$, the $H_{p}$, $B_{T}$ and $V_{T}$  passbands were adjusted until the slopes and shapes of the regressions with the synthetic photometry matched as closely as possible those of the observed regressions. In order to remove the small color term evident in the initial regressions, the  red side of the \citet{Bess00} $V_{T}$ passband was shifted slightly redward while a smaller blueward shift was made to the blue side of the \citet{Bess00} $B_{T}$ band. Fig 10 shows the final regressions for $B-B_{T}$ and $V-V_{T}$ from the NGSL spectra.  Fig 11 shows the adopted $B_{T}$ and $V_{T}$ passbands in comparison to the original passbands \citep{vLee97a} and the \citet{Bess00} passbands. There is little obvious difference between the three passbands.  The adopted photon counting response functions for $B_{T}$ and $V_{T}$ are listed in Table 2.

\begin{figure}
\figurenum{10}
\plotone{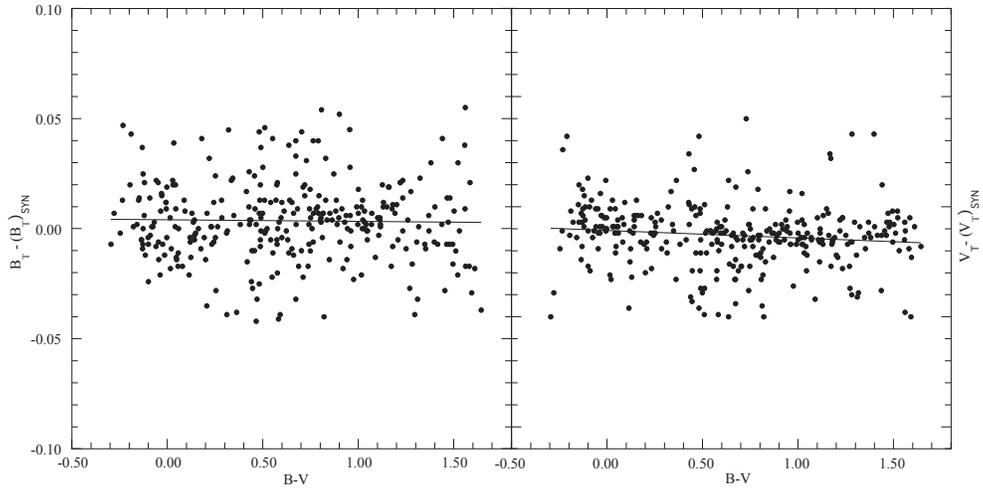}
\caption{Differences between synthetic $B_{T}$, $V_{T}$ and catalog $B_{T}$, $V_{T}$ for the NGSL sample of stars.}
\end{figure}

\begin{figure}
\figurenum{11}
\plotone{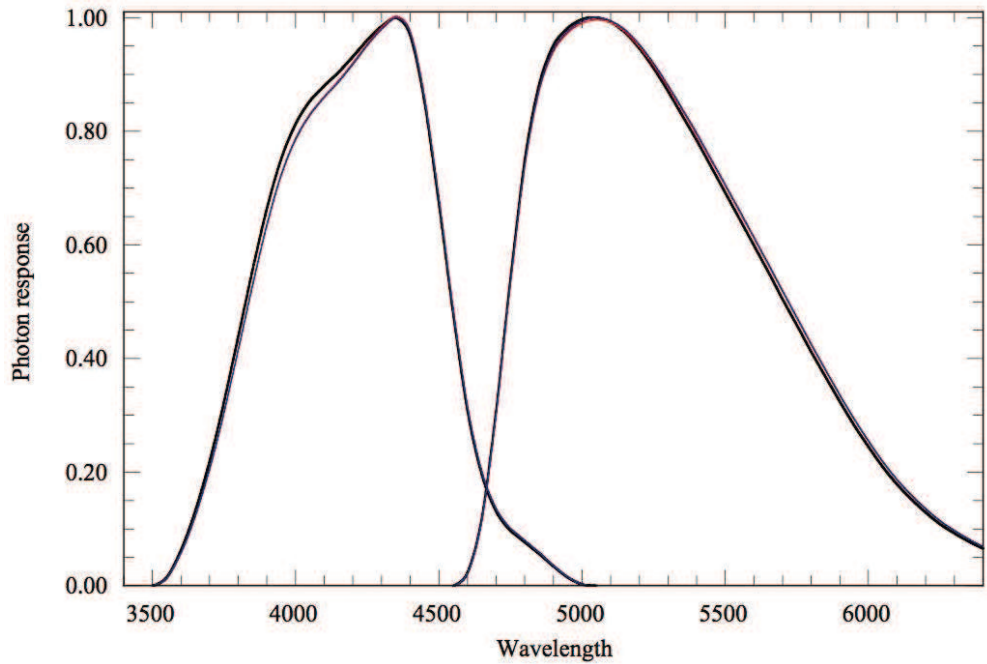}
\caption{Photon-counting response functions  $S_{x} (\lambda)$ for $B_{T}$ and $V_{T}$. This paper - thick black; \citet{vLee97a} - blue; \citet{Bes90a} - red.}
\end{figure}

There have been suggestions \citep[e.g.][]{Gren01}, that the change in the $H_{p}$ sensitivity function caused by the in-orbit radiation damage was unlikely to be a complete loss of the bluest sensitivity as suggested by \citet{Bess00}, but rather a more complicated drop in sensitivity across a wider wavelength range. We have attempted to use the two spectrophotometric samples to examine this proposition and, whilst the results are not unequivocal, a slightly better fit is achieved by making small modifications to the \citet{Bess00} passband. 

The synthetic $V-H_{p}$ versus $U-B$ , $V-I$ and  $B-V$ regressions are shown in Figs. 12, 13 and 14 respectively, in comparison with empirical relations for these stars. These plots show the range of stars represented in the NGSL spectrophotometric catalog (few K and M dwarfs but many FG subdwarfs) and the different distribution of stars in the comparison standard photometric SAAO catalogs. 
We have fitted a cubic polynomial to the $V-H_{p}$ versus $B-V$ regression for the E-region stars \citep{Menz89,Menz90} bluer than $B-V$=1.1. The same polynomial in $B-V$ was applied to the catalog stars of \citet{Pell07,Pell90} and to the synthetic photometry of the NGSL and MILES stars.

\begin{figure}
\figurenum{12}
\plotone{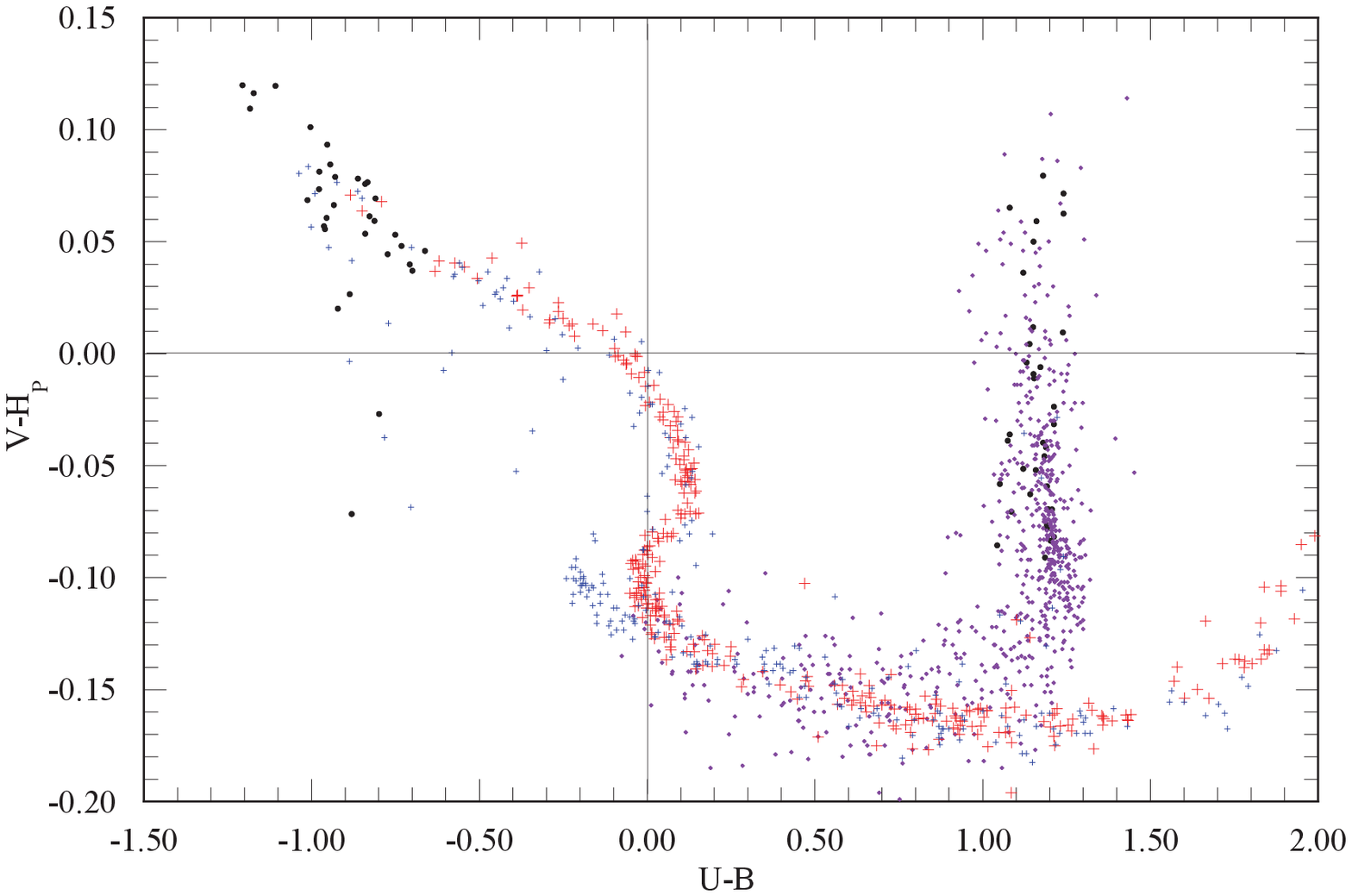}
\caption{Comparison between synthetic and catalog $V-H_{P}$ versus $U-B$ relations. Synthetic: NGSL  - blue.  Observed: E-region stars  - red; \citet{Kilk98}: more extreme O and B stars and M dwarfs - black dots; \citet{Koen10}: K and M Hipparcos dwarfs - violet. Note the large number of  metal-deficient F and G stars in the NGSL sample with $U-B$ excesses and the very few K and M dwarfs in the NGSL sample. See text for details.} 
\end{figure}

\begin{figure}
\figurenum{13}
\plotone{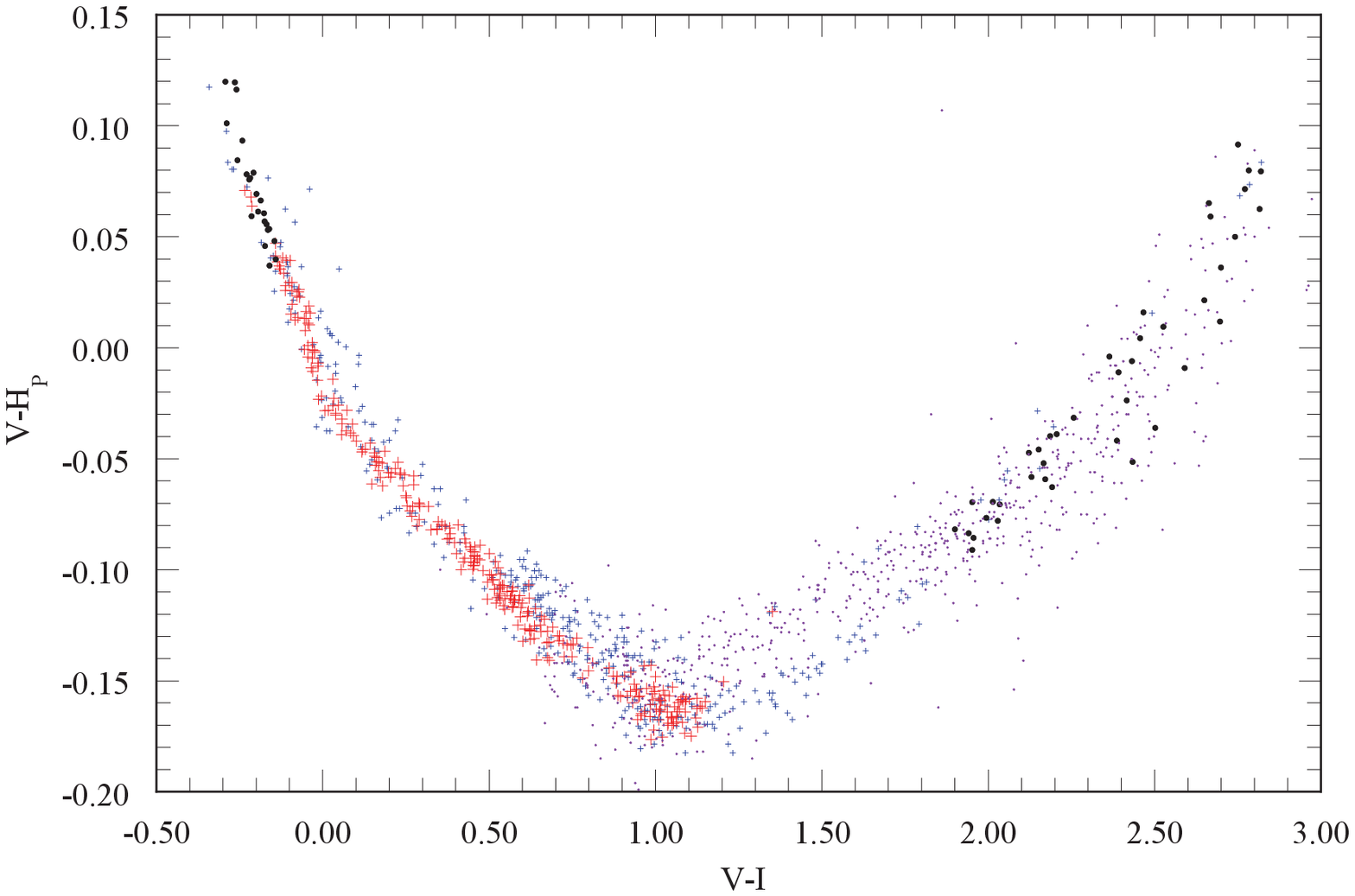}
\caption{Comparison between synthetic and catalog $V-H_{P}$ versus $V-I$  relations. Synthetic: NGSL  - blue. Observed: E-region stars  - red; \citet{Kilk98} - black;  \citet{Koen10} - violet. See text for details.} 
\end{figure}

\begin{figure}
\figurenum{14}
\plotone{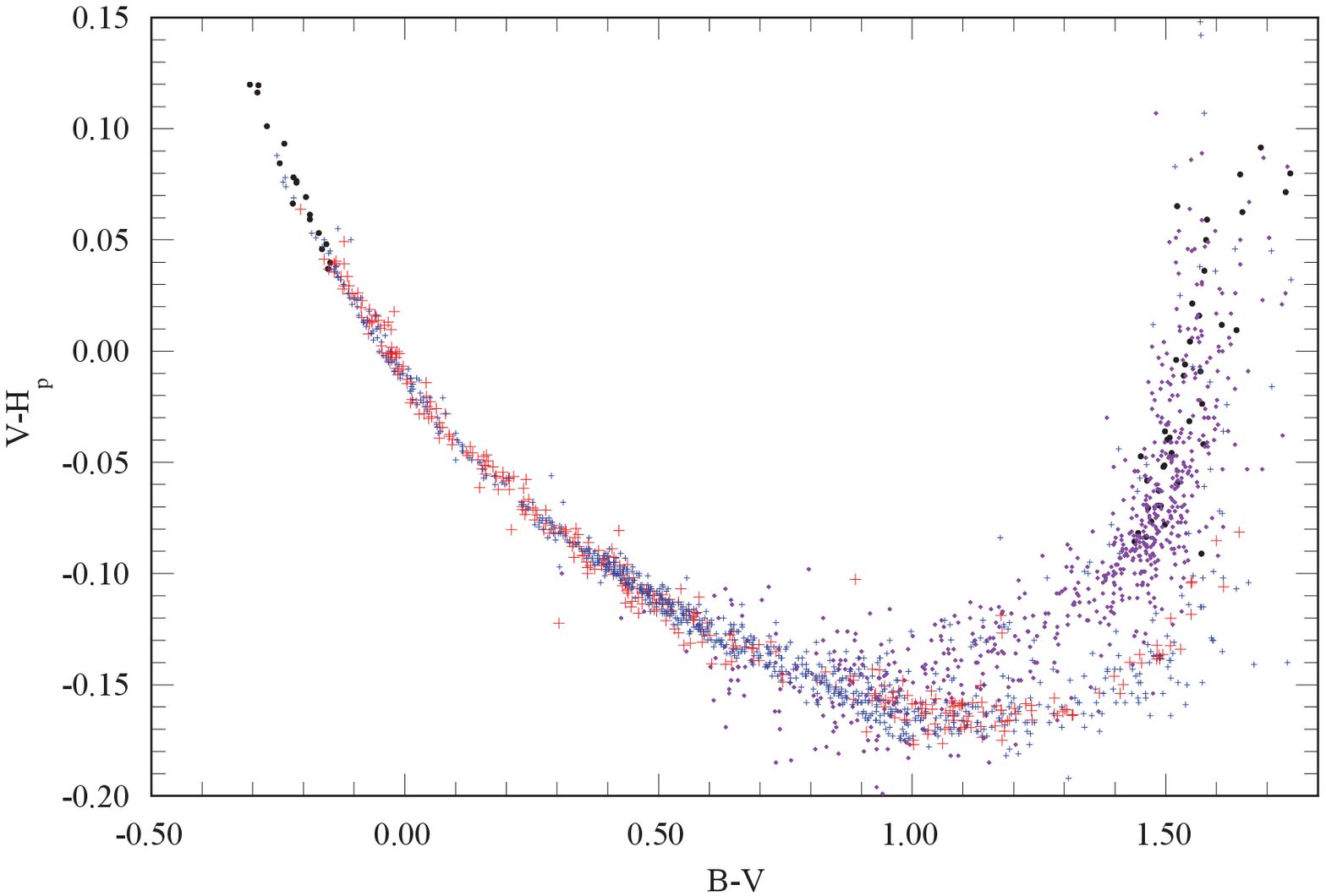}
\caption{Comparison between synthetic and catalog $V-H_{P}$ versus $B-V$ relations. Synthetic: NGSL  - blue. Observed: E-region stars - red; \citet{Kilk98} - black; \citet{Koen10} - violet. See text for details.} 
\end{figure}

In Fig 15 we plot the residuals of the fit. It is clear that the synthetic photometry using the adopted $H_{p}$ band are a very good match to the standard photometry with the caveat that the ZPs of the synthetic $V-H_{p}$ mags were adjusted to achieve this. This will be discussed in the next section.  
Table 2 also lists the new Hipparcos passband while Fig 16 shows the new and old photon counting passbands.   

\begin{figure}
\figurenum{15}
\plotone{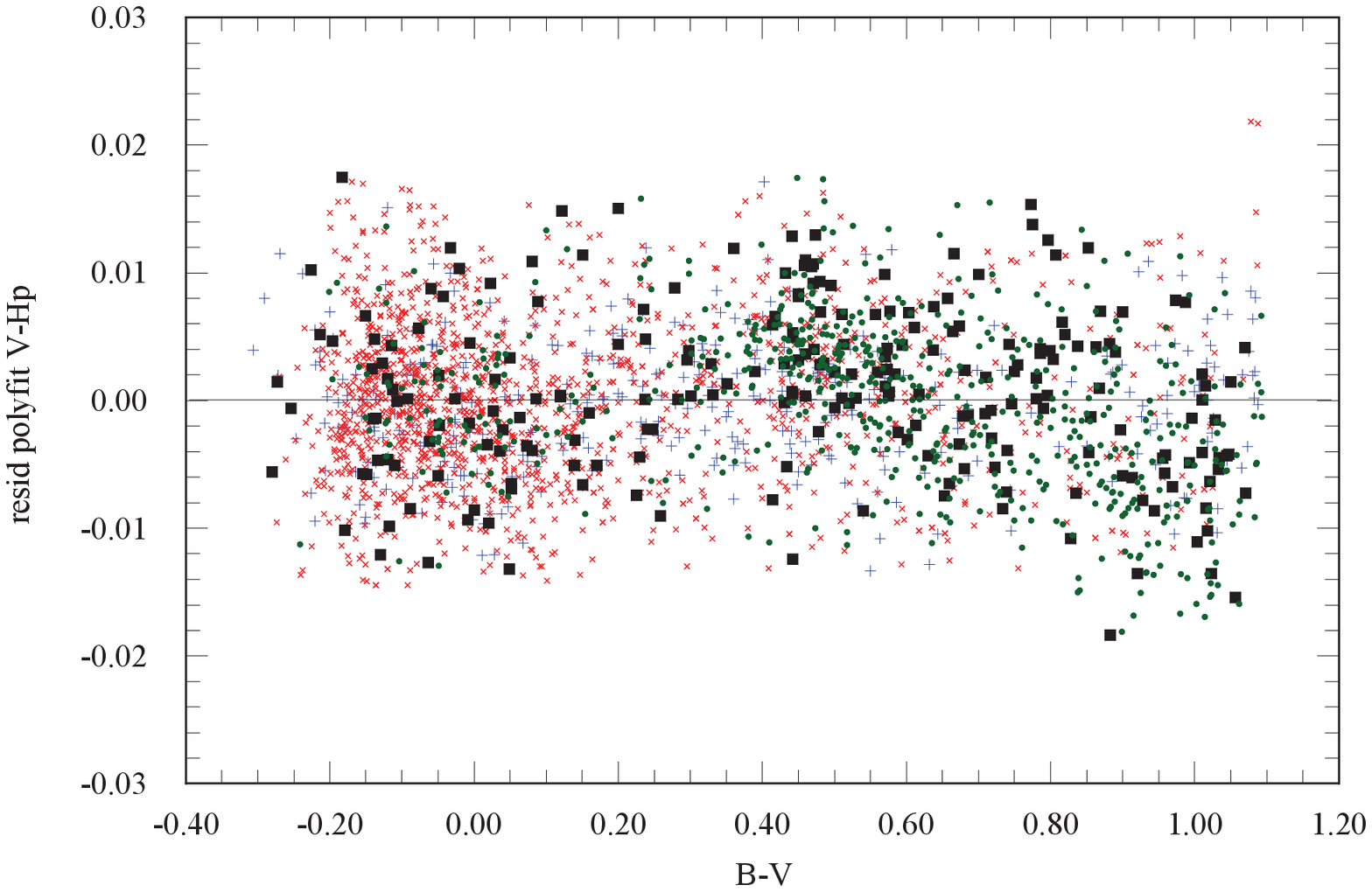}
\caption{Comparison between the residuals of the same polynomial fit to the synthetic and catalog $V-H_{P}$ versus $B-V$ relations. Synthetic: NGSL - black squares; MILES - green dots. Observed: E-region stars \citet{Menz89}, \citet{Menz90} - red crosses; \citet{Pell07}, \citet{Pell90} - blue pluses. See text for details.} 
\end{figure}

\begin{figure}
\figurenum{16}
\plotone{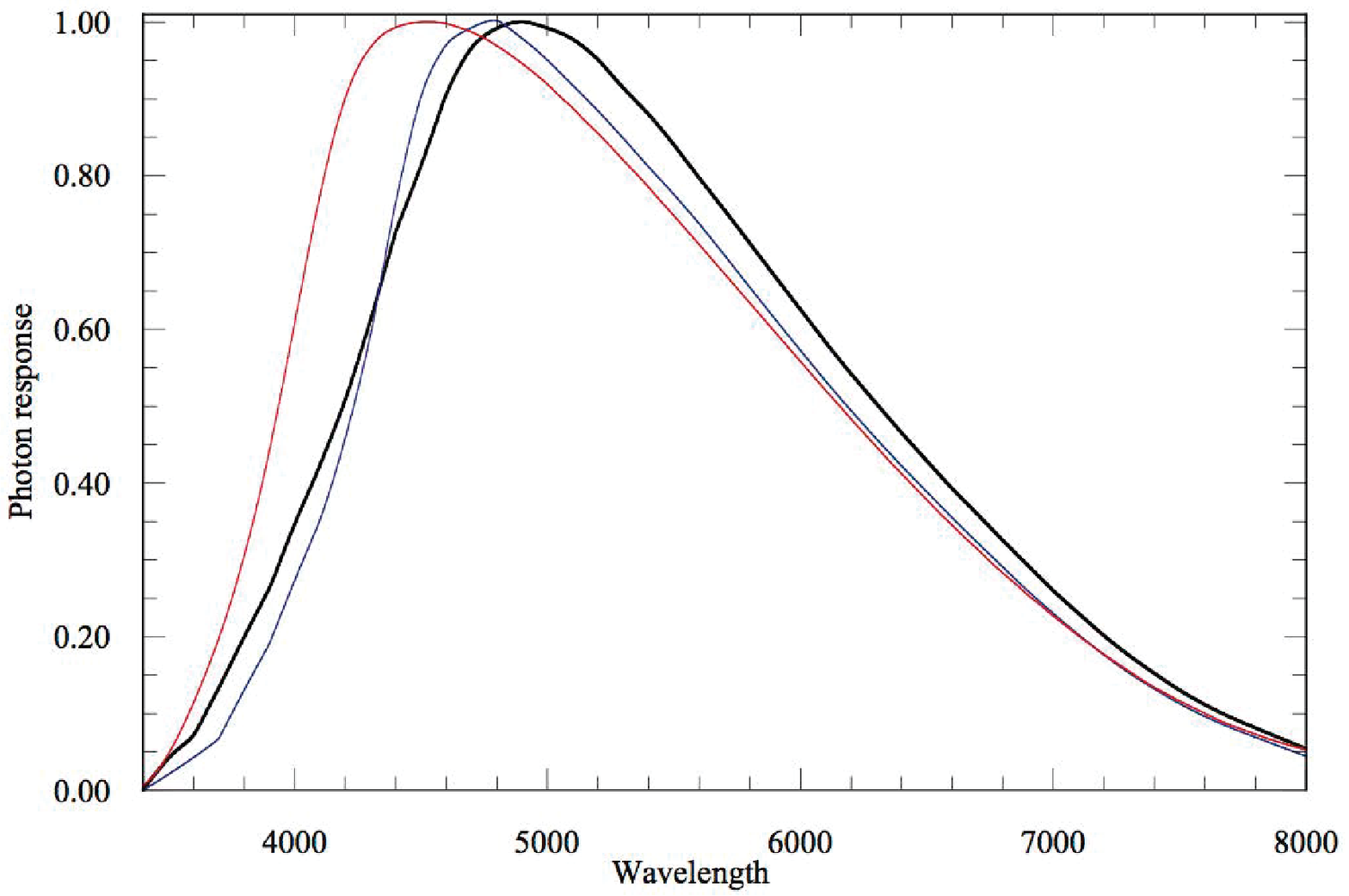}
\caption{Photonic response functions  $S (\lambda)$ for $H_{p}$. This paper - thick black; \citet{vLee97a} - red; \citet{Bes90a} - blue. See text for details.} 
\end{figure}

\section{$UBVRI$,  $H_{p}$,  $B_{T}$ and  $V_{T}$ magnitudes and zero-points}
All standard photometric systems adopt some ZP for their magnitude scale. Historically, the ZP of $V$ in the $UBV$ system is generally used for other systems. 

The AB mag system \citep{Oke83} (see Appendix) was defined as a monochromatic magnitude system for spectrophotometry where AB$_{\nu}$ = $-2.5$ log $f_{\nu}$ + 48.60 and  $f_{\nu}$ is the flux in erg cm$^{-2}$ s$^{-1}$ Hz$^{-1}$. This has now been generalized for use with  broad-band photometric bands. In the AB system, a flat spectrum star (in $f_{\nu}$) has the same AB magnitude in all passbands.  
 
The AB$_{\lambda}$ or ST mag system (see Appendix) was defined in terms of $f_{\lambda}$ where ST = $-2.5$ log $f_{\lambda}$ $+ 21.10$

The so-called VEGAMAG system (like the $UBV$ system) is one where Vega ($\alpha$ Lyrae) has colors (magnitude differences), such as $U-B$ and $B-V$, that are identically zero. This is equivalent to dividing all the observed fluxes by the flux of Vega but adjusting the $V$ ZP to give the adopted $V$ magnitude for Vega. For the Vega spectrum we used the  CALSPEC (\url{http://www.stsci.edu/hst/observatory/cdbs/calspec.html}) spectrum alpha\_lyr\_stis\_005, which is distributed in the \textit{synphot} and \textit{pysynphot} software packages (see Appendix). 

\subsection{Observed zero-points}
The Hipparcos and Tycho magnitude ZPs  \citep{vLee97b} were chosen to produce a VEGAMAG type system in which $H_{p}$ = $V_{T}$ = $V$ and $B_{T}$ = $B$ at $B-V$= 0, where $B$ and $V$ are standard magnitudes in the  Johnson-Cousins $UBV$ system.  \citet{Pell07} confirmed the excellent agreement between the $V$ magnitude scales of the homogenized $UBV$ system \citep{Nico78}, the uvby system \citep{Gron76,Olse83} the Hipparcos system and the Walraven $V, V-B$ system \citep{Pell90} (transformed with $V_{PL}$ = 6.886 $- 2.5$$V_{W}$ $-$ 0.080($V_{W}-B_{W}$). 

We have intercompared the \citet{Pell07} transformed $V_{PL}$ mags with those from the Hipparcos catalog \citep{Perr97}, the most recent homogenized $UBV$ catalog of  \citet{Merm06} and the E-region photometry from \citet{Menz89} and \citet{Menz90}.  (We also derived 
$B-V$ = 2.539($V_{W}-B_{W}$) $-$ 0.827($V_{W}-B_{W})^{2}$ + 0.3121 ($V_{W}-B_{W})^{3}$ $- 0.015$ from 1654 common stars in the \citet{Merm06} homogenized $UBV$ catalog. The transformed $B-V$ values had an rms of 0.013 mag. A slightly different fit was obtained using SIMBAD $B-V$ values.)  The results of the $V$ comparisons were $V_{Hip}=V_{PL}$ $-0.006$ (1523 stars); $V_{Merm}=V_{PL}$ $-0.005$ (1679 stars); the rms of these means are 0.0003 mag. The ZP differences are similar to the $V_{SAAO}\approx V_{PL}$ $-0.006$ reported by \citet{Cous93}. 
We derived $V-H_{p}$ for the various observed samples and by fitting a cubic to the regressions against $B-V$ for $-0.3<B-V<$1.1 have determined the  $V-H_{p}$ values for $B-V$=0. These are $-0.008$ (358 E-region stars), $-0.009$ (78 \citet{Land09} stars) and $-0.0003$ (1427 \citet{Pell07} stars). We also derived $V-V_{T}$ and $B-B_{T}$ and fitted polynomials to the $B-V$ regressions yielding $V-V_{T}$ ZPs of +0.002 (355 E- region stars)+0.008 (1618 \citet{Pell07} stars) and $B-B_{T}$ ZPs of $-0.003$ (367 E-region stars) and +0.002 (1708 \citet{Pell07} stars). 

From these comparisons  we confirm that \citet{Menz89}, \citet{Menz90}, \citet{Land09} and \citet{Merm06} have the same $V$mag ZP and  
that the transformed $V_{PL}$ mags \citep{Pell90} should be adjusted by $-0.006$ mag. (The original Walraven $VBLUW$ mags are unaffected.)  Although the ZPs of the $H_{p}$, $V_{T}$ and $B_{T}$ systems need to be adjusted by $-0.008$, +0.002 and $-0.003$ mags, respectively, to put them on the same ZP as the $UBVRI$ system,  we will retain the ZPs of the existing $H_{p}$, $B_{T}$ and $V_{T}$ systems defined by the Hipparcos and Tycho catalogs in this paper and derive synthetic photometry ZP corrections accordingly.

\subsection{Synthetic photometry zero-points}
We carried out synthetic photometry on the alpha\_lyr\_stis\_005 spectrum and assigned ZPs to force $H_{p}$= $V_{T}$= $B_{T}$ = $U$= $B$ = 0.03 (see Appendix A1.1). These Vega-based f$_{\nu}$ and f$_{\lambda}$ ZPs  are listed in Table 3.  All ZPs in this paper are to be subtracted from the AB mags (Equation 1). 

\begin{deluxetable}{lrrrrrrrr}
\tablenum{3}
\tablecaption{$UBVRI$ zero-points based on the STIS005 spectrum and $V$=0.03 for Vega }
\tablewidth{0pt}
\tablehead{\colhead{System} & \colhead{$U$} & \colhead{$B$}& \colhead{$V$}   & \colhead{$R$} & \colhead{$I$} & \colhead{$B_{T}$}& \colhead{$V_{T}$}   & \colhead{$H_{p}$}}
\startdata
AB$_{\nu}$ = abmag       &  0.771 & $-$0.138 & $-$0.023 & 0.160 & 0.402  & $-$0.090 & $-$0.044 & $-$0.022  \\
AB$_{\lambda}$ = stmag       & $-$0.142 & $-$0.625 & $-$0.019 & 0.538 & 1.220 & $-$0.672 &  $-$0.115 & $-$0.074   \\
vegamag & $-$0.023 & $-$0.023 & $-$0.023 & $-$0.023 & $-$0.023 & $-$0.023 & $-$0.023  & $-$0.023 \\
$\lambda_{eff}$  & 3673 &  4368 &  5455  & 6426  & 7939  & 4215   &  5265    & 5188     \\
\enddata
\end{deluxetable}

With these ZPs we computed synthetic photometry for all NGSL and MILES stars that had Hipparcos photometry. The ZPs from the synthetic photometry will check (a) whether there are systematic differences between the mean MILES and NGSL flux calibrations and (b) whether the STIS005 spectrum correctly represents the empirical ZPs of the $UBVRI$ and  $H_{p}$, $V_{T}$, and $B_{T}$ systems.  
We compared the synthetic magnitudes and/or colors with the observed magnitudes and colors and derived the mean differences. The few stars with exceptionally large differences were not used in the means. There were about 700 stars in the MILES sample and 300 stars in the NGSL sample. We also computed synthetic photometry for 46 of the CALSPEC spectra (http://www.stsci.edu/hst/observatory/cdbs/calspec.html), 27 of which have $UBVRI$ photometry from \citet{Land07} and \citet{Land09}, 16 had $H_{p}$ photometry and 10 had Tycho photometry. 

In Table 4 we list the mean differences. The standard errors of the means for the NGSL and MILES samples are less than 0.001 mag. There was good agreement between the NGSL and MILES $VR$ $I$, $V_{T}$ and $H_{p}$ results; however, the differences for $B$ and $B_{T}$ appear to be small but real. We chose to adopt the NGSL values for $B$ and $B_{T}$ in preference to the MILES values as the NGSL data were taken outside the atmosphere and are unaffected by atmospheric extinction. 
For the far fewer CALSPEC spectra, the errors in the mean were between 0.004 ($H_{p}$) and 0.04 ($B_{T}$ and $V_{T}$). Given the small number of CALSPEC stars with photometry, the mean differences in the colors of the much fainter CALSPEC spectra were in reasonable agreement with those for the NGSL sample, except for an unexplained difference of a few hundredths of a magnitude between the $V$ and H$_{p}$ magnitudes. 

\begin{deluxetable}{lrrrrrrrrr}
\tablenum{4}
\tablecaption{Mean differences between the synthetic and observed photometry}
\tablewidth{0pt}
\tablehead{\colhead{Source} & \colhead{$V$} & \colhead{$U-B$}& \colhead{$B-V$}   & \colhead{$V-R$} & \colhead{$V-I$} & \colhead{$B_{T}$}& \colhead{$V_{T}$}   & \colhead{$V-H_{p}$} & N$_{stars}$ }
\startdata
NGSL & 0.002 & $-$0.018 & 0.005 &$\sim$0.0    & 0.004  & $-$0.010  & 0.007 &  0.012 & $\sim$300 \\
MILES & 0.004 &            &  $-$0.012 & $\sim$0.0 & $-$0.006 &  $-$0.032  & 0.007 & 0.012 & $\sim$700  \\
\enddata
\end{deluxetable}

In Table 5 we list the additional ZP mag offsets that will place synthetic photometry computed with the AB mag ZPs from Table 3 on the same scale as the  homogeneous $UBV$ system, the Cousins-Landolt $UBVRI$ system and the Hipparcos and Tycho systems. We also list two wavelengths associated with each passband, that are defined independently of the flux; the pivot wavelength, $\lambda{_p}$, and the mean photon wavelength $\lambda{_0}$ (see Appendix for details) and the FWHM of the passband; the wavelengths are given in \AA. Note that these are the wavelengths that should be associated with published $UBVRI$ photometry, not the natural passbands used by various observers, as their photometry has been transformed onto the standard system.

\begin{deluxetable}{lrrrrrrrr}\tt
\tablenum{5}
\tablecaption{Adopted additional\tablenotemark{1} ZP magnitudes and passband parameters}
\tablewidth{0pt}
\tablehead{\colhead{  } & \colhead{$U$} & \colhead{$B$}& \colhead{$V$}   & \colhead{$R$} & \colhead{$I$} & \colhead{$B_{T}$}& \colhead{$V_{T}$}   & \colhead{$H_{p}$} }
\startdata
ZP & $-$0.010 & 0.008 & 0.003&0.003& 0.002 & $-$0.010  &  0.007 & $-$0.008\\
$\Delta$$\lambda_{fwhm}$  & 625 &  890 &  830  & 1443  & 1499  & 718   &  962    & 2116 \\
$\lambda_{p}$  & 3597 &  4377 &  5488  & 6515  & 7981  & 4190   &  5300    & 5347 \\
$\lambda_{0}$  & 3603 &  4341 &  5499  & 6543  & 7994  & 4198   &  5315    & 5427 \\
\enddata
\tablenotetext{1}{These zero-points are to be applied in addition to those in Table 3 based on the stis005 spectrum of Vega.}
\end{deluxetable}

The uncertainties in these additional zero-points should only be a few milli-mags, except for $R$ and $I$ where it is more uncertain, as the available $V-I$ photometry was of lower precision. 

Using these total ZP corrections we recomputed the mags for the Vega stis005 spectrum and obtained $V$ = 0.027, $U-B$ = 0.018, $B-V = -0.004$, $V-R$ = 0.000, $V-I = -0.001$. In addition, the 1994 ATLAS Vega spectrum of Castelli (\url{ http://wwwuser.oat.ts.astro.it/castelli/vega/fm05t9550g395k2odfnew.dat}) gives $U-B = -0.017$, $B-V = -0.017$, $V-R = -0.004$, $V-I = -0.009$. For comparison, \citep{Bess83} measured for Vega, $B-V = -0.01$, $V-R= -0.009$, $V-I = -0.005$. 

\section{Summary}
Excellent spectrophotometric catalogs are now available from NGSL \citep[][\url{http://archive.stsci.edu/prepds/stisngsl/index.html}]{Heap07}  and MILES \citep[][\url{http://www.iac.es/proyecto/miles/}]{Sanc06}. In addition to their intrinsic worth, such stars are very useful to use to calibrate all-sky surveys, such as SkyMapper \citep{Kell07}. However, the published absolute flux levels are imprecise or non-existent, so we have renormalized the spectra to their precise $H_{p}$ magnitudes. In order to do this it was necessary to determine the best $H_{p}$ passband. We also decided to reexamine  the passbands representing the $UBVRI$ and Tycho $B_{T}$ and $V_{T}$ standard photometric systems using the renormalized NGSL and MILES spectra. We used the CALSPEC stis005 spectrum of Vega to derive the nominal ZP corrections to the AB mag fluxes and synthesized the various magnitudes and slightly adjusted the photonic passbands, achieving better agreement between the synthetic and standard magnitudes than possible with the previous  passbands. In Table 1 and 2 we presented the adopted photonic passbands for $UBVRI$ and $H_{p}$, $B_{T}$ and $V_{T}$, respectively. Table 3 lists the ZP magnitude corrections based on the stis005 Vega spectrum and $V$ = 0.03.

We intercompared the ZPs of the $V$ magnitude scale of the SAAO Cousins-Landolt $UBVRI$ system, the homogenized \citet{Merm06} $UBV$ system and the Walraven \citet{Pell07} system and the Hipparcos $H_{p}$ and Tycho $B_{T}$ and $V_{T}$ systems. We found small differences of less than 0.01 mag between them.  The $H_{p}$ magnitude ZP differs by 0.008 mag from the ZP of the $UBVRI$ system.  

We analysed the mean magnitude and color differences between the synthetic photometry and the standard photometry and proposed small additional ZP corrections to place the synthetic photometry computed using the AB mag ZPs in Table 3 onto the same ZPs as the standard system photometry. These additional ZP corrections are given in Table 5 together with the passband parameters that should be used to characterize the standard systems. 

There was good agreement between the mean differences from the NGSL and MILES catalogs, although the mean level of the MILES blue fluxes deviated slightly, but systematically, from those of the NGSL spectra. The synthetic colors of the fainter CALSPEC spectra also supports the proposed additional ZP corrections except for an unexplained difference in the relative $V$ and $H_{p}$ magnitudes.   

Finally, in  the Appendix, we present an extensive discussion on the confusion in the literature concerning measured magnitudes, fluxes and response functions when broad-band photometry is involved and provided equations setting out clearly the derivation of photometric  quantities.  We also cross reference parameters and definitions used in the HST photometric packages \textit{synphot} and its successor \textit{pysynphot}. (Table A1). 

 \acknowledgments
We wish to thank Sally Heap for correspondence concerning the re-reduction of the NGSL spectra; Jan Willem Pel and Jan Lub for a digital version of their Walraven photometry catalog and for helpful discussions; and Wolfgang Kerzendorf for fitting and extrapolating the MILES spectra to cover the 3000\AA\--10200\AA\ wavelength region. We thank the referee for suggestions to make the paper more accessible to physicists. Vizier-R, Simbad, TOPCAT and Kaleidagraph were used in preparing this paper.  

\appendix

\section{APPENDIX}
There is unfortunately some confusion in the literature concerning measured magnitudes, fluxes and response functions when broad-band photometry is discussed. The definitions concerning monochromatic fluxes are clear -- but see \citet{Soff99} concerning the paradoxes, errors, and confusions that arise when density distributions are involved --  but the clarity is lost when these definitions are generalized to involve mean magnitudes, mean fluxes and the choice of the ``effective"  wavelength or frequency most appropriately associated with them. 
\subsection{Photometric quantities and definitions}
In astronomy, flux ($f$) refers to the radiative flux density, a quantity in physics referred to as the spectral irradiance. In astronomy, flux is also referred to as the monochromatic flux $f_{\nu}$ or $f_{\lambda}$, to distinguish it from the total flux $F$ which is summed over all wavelengths or frequencies.  In SI units, $f_{\lambda}$ is measured in W m$^{-3}$, or more practically in W m$^{-2}$ \AA$^{-1}$, W m$^{-2}$ nm$^{-1}$ or W m$^{-2}$ $\mu$m$^{-1}$, depending on the part of the spectrum being considered. In cgs units it is measured in  erg cm$^{-2}$ sec$^{-1}$ \AA$^{-1}$, erg cm$^{-2}$ sec$^{-1}$  nm$^{-1}$ or erg cm$^{-2}$ sec$^{-1}$ $\mu$m$^{-1}$. (10$^{3}$ erg  cm$^{-2}$ sec$^{-1}$  = 1 W m$^{-2}$). In radio-astronomy, fluxes are usually expressed in terms of a non-SI unit, the flux unit or Jansky (Jy), which is equivalent to 10$^{-26}$ W m$^{-2}$ Hz$^{-1}$ or 10$^{-23}$ erg cm$^{-2}$ sec$^{-1}$ Hz$^{-1}$.

A good starting point for the relevant formulae and definitions used in photometry is  \citet{Rufe88}, \citet{Koor86} and \citet{Toku05}.
The stellar flux is normally given in terms of $f_{\nu}$ or $f_{\lambda}$, and the units, respectively, are 
erg cm$^{-2}$ sec$^{-1}$ Hz$^{-1}$ and erg cm$^{-2}$ sec$^{-1}$ \AA $^{-1}$ or in the SI system of units, W m$^{-2}$ Hz$^{-1}$ and W m$^{-2}$ nm$^{-1}$; although, rather than energy, the photon flux $n_{p}$ in photon m$^{-2}$ sec$^{-1}$ Hz$^{-1}$ or photon m$^{-2}$ sec$^{-1}$ \AA $^{-1}$ is also used. 
The relations between these quantities are precisely defined for monochromatic light, namely    \
\begin{equation}
$$ $f_{\nu}$ = f$_{\lambda}$$\dfrac{\lambda^{2}} {c}$ $$
\end{equation}
and
\begin{equation}
$$ $n_{p}$ = $f_{\lambda}$$\dfrac{\lambda} {h c}$ $$
\end{equation}

The AB (absolute) magnitude scale was introduced by \citet{Oke65} who proposed the $f_{\nu}$ definition, having noted that a plot of  $f_{\nu}$ versus 1/$\lambda$ for hot stars, was approximately linear in the optical part of the spectrum. The monochromatic magnitude AB was later defined by \citet{Oke83} using the flux measurement adopted by \citet{Oke70} for Vega at 5480\AA\  and an apparent magnitude of $V$ = +0.035. The Vega flux was considered measured to an accuracy of about 2 percent. \citet{Oke70} measured the flux of Vega at a set of discrete 50\AA\ bands.  A mean value of $f_{\nu}$ = 3.46 $\times$ 10$^{-20}$ erg cm$^{-2}$ sec$^{-1}$ Hz$^{-1}$  or 3.36 $\times$ 10$^{-9}$ erg cm$^{-2}$ sec$^{-1}$ \AA $^{-1}$ or 940  photon cm$^{-2}$ sec$^{-1}$ \AA $^{-1}$ was measured at 5556\AA. They then interpolated Vega's flux to the value of 3.65 $\times$  10$^{-20}$ erg cm$^{-2}$ sec$^{-1}$ Hz$^{-1}$ at 5480\AA\ assumed to be the ``effective" wavelength of the $V$ band and using this value together with $V$ = +0.035, derived the constant $-$48.60 associated with definition for the AB magnitude, namely
\begin{equation}
$$ AB$_{\nu}$ = $-$2.5 log$f_{\nu}$ $-$ 48.60 $$
\end{equation}
It is somewhat unfortunate that Oke chose to define the AB mag in terms of $f_{\nu}$ rather than $f_{\lambda}$, which is more appropriate for most stars; but the conversions, at least for monochromatic light,  are straightforward.
\begin{equation}
$$ AB$_{\lambda}$ = $-$2.5 log$f_{\lambda}$ $-$ 21.10 $$
\end{equation}
AB$_{\lambda}$ is called STMAG in \textit{synphot} and \textit{pysynphot}. Note that these ZPs are based on the nominal wavelength of 5480\AA\ for the $V$ band. 
 
More recent measurements of Vega's flux are about 2 percent brighter, and retaining the above values of the ZPs in the definition of AB mag and ST mag will mean that these scales will necessarily have different ZPs from the $V$ system. And if a different nominal wavelength for the $V$ band is adopted, this will introduce an additional systematic difference between the $f_{\nu}$  and $f_{\lambda}$ ZPs. 

\subsubsection{The flux and $V$ mag of Vega}
Summaries of the direct measurements of the optical flux of Vega are given by \citet{Hay85} and \citet{Mege95}, who proposed $f_{\lambda}$ = 3.44 $\pm$ 0.05  $\times$  10$^{-9}$ and 3.46 $\pm$ 0.01 $\times$ 10$^{-9}$ erg cm$^{-2}$ sec$^{-1}$ \AA $^{-1}$, respectively, for Vega at 5556\AA. \citet{Coh92} adopted the Hayes value, together with the flux spectrum of a Vega ATLAS 9 model for their spectral irradiance calibration.  More recently, \citet{Bohl04} measured the flux for Vega using STIS spectra and \citet{Bohl07} refined these observations and discussed model fits, including rapidly rotating pole-on models. \citet{Bohl07} quoted an absolute flux at 5556\AA\  the same as \citet{Mege95} and $V$ = 0.023 and adopted for Vega a combination of various source fluxes to produce the  CALSPEC spectrum alpha\_lyr\_stis\_005 that is now generally used by \textit{pysynphot} and other routines. 

Many direct $V$ measurements of Vega have been made over the years. An obvious problem has been its extreme brightness making it difficult to measure with sensitive photomultipliers on 1m telescopes; however, \citet{Bess83} measured $V$ = 0.03 in comparison with Cousins bright equatorial stars using an Inconel coated 1\% neutral density filter and a GaAs photomultiplier tube at Kitt Peak. This value is in exact agreement with \citet{John66}. \citet{Hay85} discussed measurements of the $V$ magnitude of Vega and discounted reports of its variability. More recently \citet{Gray07} also discussed observations of Vega.  \citet{Merm06} gives for Vega $V$ = 0.033 $\pm$0.012. 

We have computed $V$ = 0.007 from the CALSPEC spectrum of Vega using our passband and the ZP of $-$48.60 in equation A3. This implies that ZPs of $-$48.58 and $-$21.08, respectively, would put the AB$_{\nu}$ and  AB$_{\lambda}$ mag scale on the same ZP as the $V$ mag system, but see Section 7.2 above.
 
\subsection{Issues arising from a broad passband} 
Photometric observations are normally made by summing the flux over discrete wavelength intervals defined by a window (filter) function. A generalized filter function is a dimensionless (unitless) quantity $R$ representing the fraction of the flux $f$ at each wavelength, that is incident on the detector. It is the product of the atmospheric transmission, the mirror reflectivity, the optics transmission and the glass filter transmission.  It is usually used in the form of a normalized function. The mean $f_{\lambda}$ flux would be expressed by the following equation
\begin{equation}
$$ $\langle f_{\lambda} \rangle$  =  $ \frac{\bigint{f_{\lambda}(\lambda) R(\lambda) d\lambda}} {\bigint{R(\lambda) d\lambda}} $  $$
\end{equation}
and a similar equation for $\langle f_{\nu} \rangle$ with all $\lambda$ replaced by $\nu$. All integrals are nominally from zero to infinity but are sensibly evaluated over the defined range of the filter passband.   

One source of confusion is the fact that the flux is evaluated after the detector, not before it. This means that the filter function must be multiplied by the response function of the detector to give the system response function, as the detector converts the incident light into electrons, which are then amplified and measured. In the case of a photon counting detector, such as a CCD, the function $R$ is multiplied by the quantum efficiency $\eta$$(\lambda)$ of the CCD to give the system photon response function $S$. In the case of detectors with photocathodes using D.C. techniques and current integration, the function $R$ is multiplied by the photocathode radiant response $\sigma(\lambda)$  (in units of mA/W) to give the system energy response function $S'$.   Incorrect equations for photon counting and energy integration  \citep[e.g.][equation 1 and 2] {Buse86} result from overlooking this difference. The system response functions are then generally renormalized.

The relations between $S(\lambda), S'(\lambda)$, $\eta(\lambda)$ and $\sigma(\lambda)$ are
\begin{equation} 
 S(\lambda) =   R(\lambda)\eta(\lambda)  \quad \textrm{and} \quad S'(\lambda) =   R(\lambda)\sigma(\lambda) 
\end{equation}
\begin{equation} 
$$  $\eta$$(\lambda)$ = $\dfrac{h c} {e \lambda}\sigma(\lambda)$  = $\dfrac{12.4}{\lambda}\sigma(\lambda)$, for $\lambda$ in \AA\ , $\eta$ in percent and $\sigma$ in mA W$^{-1}$ $$
\end{equation}
or
\begin{equation} 
$$  $\eta(\nu)$ = $\dfrac{h \nu} {e}\sigma_{\nu}(\nu)$  $$
\end{equation}

Ignoring the atomic constants for the moment (and noting that the response functions are usually normalized) we can write
\begin{equation} 
$$  $S'(\lambda)$ =   $R(\lambda)\eta(\lambda)\lambda$ = $S(\lambda) \lambda$ $$
\end{equation}
and
\begin{equation} 
$$ $S'(\nu)$  =   $R(\nu)\eta(\nu)/\nu$=  $S(\nu)/\nu$ $$
\end{equation}
So for an energy integrating detector we can express the measured mean energy  flux as 
\begin{equation} 
$$ $\langle f_{\lambda} \rangle$  =   $ \frac{\bigint{f_{\lambda}(\lambda) S'(\lambda) d\lambda}} {\bigint{S'(\lambda) d\lambda}} $  =  $ \frac{\bigint{f_{\lambda}(\lambda) S(\lambda) \lambda d\lambda}} {\bigint{S(\lambda) \lambda d\lambda}} $   $$  
\end{equation}
and for  $\langle f_{\nu} \rangle$ 
\begin{equation} 
$$ $\langle f_{\nu} \rangle$  =   $\frac{\bigint{f_{\nu}(\nu) S'(\nu)  d\nu}} {\bigint{S'(\nu) d\nu}} $ = $\frac{\bigint{f_{\nu}(\nu) S(\nu)  d\nu/\nu}} {\bigint{S(\nu) d\nu/\nu}} $=  $\frac{\bigint{f_{\lambda}(\lambda) \dfrac{\lambda^{2}} {c} S(\lambda) \lambda  \dfrac{c} {\lambda^{2}} d\lambda}} {\bigint{S(\lambda) \lambda  \dfrac{c} {\lambda^{2}}d\lambda}} $  =  $\frac{\bigint{f_{\lambda}(\lambda) S(\lambda) \lambda d\lambda}} { \bigint{S(\lambda) \dfrac{c} {\lambda}d\lambda}} $      $$
\end{equation}

For photon-counting detectors, one electron is collected for every detected photon so the mean photon flux is given by 
\begin{equation}
$$ $\langle n_{p} \rangle$  =  $\frac{\bigint{n_{p}(\lambda) S(\lambda) d\lambda}} {\bigint{S(\lambda) d\lambda}} $   
= $\frac{\bigint{f_{\lambda}(\lambda) \dfrac{\lambda} {h c} S(\lambda) d\lambda}} {\bigint{ \dfrac{\lambda} {h c} S(\lambda)d\lambda}}$ $\times$ 
$\frac{\bigint{S(\lambda)  \dfrac{\lambda} {h c} d\lambda}}  {\bigint{S(\lambda) d\lambda}}$ = $\langle f_{\lambda} \rangle$ $\dfrac{\lambda_{0}}{h c}$  $$  
\end{equation}
where 
\begin{equation}
$$ $\lambda_{0}$ =  $\frac{\bigint{\lambda S(\lambda)  d\lambda}} {\bigint{S(\lambda) d\lambda}}$ $$
\end{equation}
Equation A13 is very important because it shows that in broad band photometry, the mean photon flux is proportional to the mean energy flux and counting photons is equivalent to integrating the energy. Furthermore, the wavelength $\lambda_{0}$  is the representative wavelength of the mean photons and could be called the mean photon wavelength of the passband.  

\subsubsection{Definitions of other wavelengths and frequencies associated with a passband}
Now, because monochromatically $f_{\nu}$ = f$_{\lambda} \dfrac{\lambda^{2}} {c}$ we can write that 
\begin{equation}
$$ $\langle f_{\nu} \rangle$ = $\langle f_{\lambda} \rangle$$\dfrac{\lambda_{p}^{2}} {c}$  $$
\end{equation}
where  $\lambda_{p}$ is called the pivot-wavelength and from equations A11 and A12 can be shown to be
\begin{equation}
$$ $\lambda_{p}$ =  $\sqrt{\frac{\bigint{S(\lambda) \lambda d\lambda}} {\bigint{\dfrac{S(\lambda)} { \lambda} d\lambda}}}$ $$
\end{equation}
As noted by \citet{Koor86}, the pivot-wavelength is convenient because it allows an exact conversion between the mean broadband fluxes
$\langle f_{\nu} \rangle$ and $\langle f_{\lambda} \rangle$. 

We defined above the mean photon wavelength $\lambda_{0}$; we could also define the mean energy wavelength that is 
\begin{equation}
$$ $\lambda'_{0}$ =  $\frac{\bigint{\lambda S'(\lambda) d\lambda}} {\bigint{S'(\lambda) d\lambda}}$  =  $\frac{\bigint{S(\lambda) \lambda^{2} d\lambda}} {\bigint{S(\lambda) \lambda d\lambda}}$ $$
\end{equation}
This mean wavelength was discussed by \citet{King52} who cites it as being favoured by Stromgren and Wesselink as a flux independent ``effective" wavelength for broad band systems. 

Two other wavelengths are commonly used: the isophotal wavelength  and the effective wavelength. 

The isophotal wavelength $\lambda_{iso}$, recommended by \citet{Coh92}, is the wavelength at which the interpolated, smoothed monochromatic flux has the same value as the mean flux integrated across the band. That is
\begin{equation}
$$ $f_{\lambda}(\lambda_{iso})$  =  $\langle f_{\lambda} \rangle$ =  $\frac{\bigint{f_{\lambda}(\lambda) S'(\lambda) d\lambda}} {\bigint{S'(\lambda) d\lambda}}$ = $\frac{\bigint{f_{\lambda}(\lambda) S(\lambda) \lambda d\lambda}} {\bigint{S(\lambda) \lambda d\lambda}}$   $$  
\end{equation}
A similar expression can be written for the isophotal frequency.  
\begin{equation}
$$ $f_{\nu}(\nu_{iso})$  =  $\langle f_{\nu} \rangle$ = $\frac{\bigint{f_{\nu}(\nu) S(\nu)/\nu d\nu}} {\bigint{S(\nu)/\nu d\nu}}$   $$  
\end{equation}
Note that both definitions relate to the energy flux but a different pair of equations could be defined in terms of the photon flux. 
\begin{equation}
$$ $ n_{p}(\lambda_{iso})$  =  $\langle n_{p} \rangle$ =  $\frac{\bigint{n_{p}(\lambda) S(\lambda) d\lambda}} {\bigint{S(\lambda) d\lambda}}$   =  $\dfrac{(1/hc)\bigint{f_{\lambda}(\lambda) S(\lambda) \lambda d\lambda}} {\bigint{S(\lambda) \lambda d\lambda}}$ $\times$ $\frac {\bigint{S(\lambda) \lambda d\lambda}}{\bigint{S(\lambda) d\lambda}}$ =  $\langle f_{\lambda} \rangle$ $\dfrac{\lambda_{0}}{h c}$ $$  
\end{equation}

The effective wavelength is usually defined as the flux-weighted mean wavelength.
In terms of photons: 
\begin{equation}
$$ $\lambda_{eff}$  =  $\frac{ \bigint{\lambda n_{p}(\lambda) S(\lambda) d\lambda}} {\bigint{n_{p}(\lambda) S(\lambda) d\lambda}} $ = $\frac{ \bigint{\lambda^{2} f_{\lambda}(\lambda) S(\lambda) d\lambda}} {\bigint{\lambda f_{\lambda}(\lambda) S(\lambda) d\lambda}} $  $$  
\end{equation}
which is the same as in terms of energy:
\begin{equation}
$$ $\lambda'_{eff}$  =  $ \frac{\bigint{\lambda f_{\lambda}(\lambda) S'(\lambda) d\lambda}} {\bigint{ f_{\lambda}(\lambda) S'(\lambda) d\lambda}} $  =  $ \frac{\bigint{\lambda^{2} f_{\lambda}(\lambda) S(\lambda) d\lambda}} {\bigint{ f_{\lambda}(\lambda) S(\lambda) \lambda d\lambda}} $ $$  
\end{equation}
and
\begin{equation}
$$ $\nu'_{eff}$  =  $\frac{\bigint{\nu f_{\nu}(\nu) S'(\nu)d\nu}} {\bigint{f_{\nu}(\nu) S'(\nu) d\nu}} $  =  $ \frac{\bigint{f_{\nu}(\nu) S(\nu) d\nu}} {\bigint{f_{\nu}(\nu) S(\nu) d\nu/\nu}} $ $$  
\end{equation}

Note that $\lambda_{iso}$ and $\lambda_{eff}$ depend explicitly on the underlying flux distribution through the filter. 

The definitions and labels of the various wavelengths and frequencies have long stirred passions. \citet{King52} argued strongly against the currently accepted use of ``effective wavelength" and ``effective frequency"  as defined in equations A22, A23 noting that the meaning of ``effective wavelength" is  better served by the isophotal wavelength. He further proposed that the mean wavelengths, defined in equations A14 and A17  to a first approximation, act as  effective wavelengths for all stars, being independent of the flux distribution. 

Finally, \citet{Schn83} for aesthetic reasons, defined the effective frequency for a passband to be 
\begin{equation}
$$  $\nu^{*}_{eff}$  =   exp$\langle \text{ln } \nu \rangle$ = exp$\frac{ \bigint{ (\text{ln } \nu) n_{p}(\nu) S(\nu) d\nu}} {\bigint{n_{p}(\nu) S(\nu) d\nu}}$   $$ 
\end{equation}
and it follows that 
\begin{equation}
$$  $\lambda^{*}_{eff}$  =   exp$\langle \text{ln } \lambda \rangle$ = exp$\frac{ \bigint{ (\text{ln } \lambda) n_{p}(\nu) S(\nu) d\nu}} {\bigint{n_{p}(\nu) S(\nu) d\nu}}$ = exp$\frac{ \bigint{ (\text{ln } \lambda) f_{\nu}(\nu) S(\nu) (1/\nu)d\nu}} {\bigint{f_{\nu}(\nu) S(\nu) (1/\nu)d\nu}}$

 =  exp$\frac{ \bigint{ (\text{ln } \lambda) f_{\nu}(\nu) S(\nu) d(\text{ln } \nu)}} {\bigint{f_{\nu}(\nu) S(\nu) d(\text{ln } \nu)}}$ $$ 
\end{equation}
We note that this is not what \citet{Fuku96} claimed \citet{Schn83} defined as $\lambda_{eff}$.  
\citet{Fuku96} defined 
\begin{equation}
$$  $\lambda_{eff}$  =   exp$\langle \text{ln } \lambda \rangle$ = exp$\frac{ \bigint{ (\text{ln } \lambda) S(\nu) d(\text{ln }\nu)}} {\bigint{S(\nu) d(\text{ln }\nu)}}$   $$ 
\end{equation}
which is not flux averaged as is the $\lambda_{eff}$ of \citet{Schn83}. 
To further complicate definitions, \citet{Doi10} defined 
\begin{equation}
$$  $\lambda_{eff}$  =   $\dfrac{c}{\nu_{eff}}$ $$ 
\end{equation}
where 
\begin{equation}
$$  $\nu_{eff}$  =   $\frac{ \bigint{\nu S(\nu) d\nu/h\nu}} {\bigint{S(\nu) d\nu/h\nu}}$ = $\frac{ \bigint{\nu S'(\nu) d\nu}} {\bigint{S'(\nu) d\nu}}$  $$ 
\end{equation}
Equation A28 is the definition of the mean energy-weighted frequency not the mean (effective [sic]) photon weighted frequency as stated by \citet{Doi10}. Footnote 13 in that paper is also in error. 

The ``effective" frequency defined by \citet{Doi10} is the ``mean" frequency defined by \citet{Koor86} and not the usual definition of the flux-weighted ``effective" frequency similar to the \citet{Schn83} ``effective" frequency. 

For comparison we evaluate the labelled wavelengths for the $V$ band and Vega. Some, such as $\lambda_{eff}$,  involve the product of the stellar flux and the system response function, while others such as $\lambda_{p}$ and $\lambda_{0}$ concern the system response passband only. Some of the different labelled photon counting wavelength values are ($\lambda_{eff}, \lambda_{iso}, \lambda_{p}$) = (5455\AA, 5486\AA, 5488\AA). The mean photon wavelength $\lambda_{0}$ = 5499\AA\ compared with the mean energy wavelength $\lambda'_{0}$ = 5524\AA. The \citet{Schn83} $\lambda^{*}_{eff}$ = 5444\AA, the \citet{Fuku96} $\lambda_{eff}$ = 5464\AA\ and the \citet{Doi10} $\lambda_{eff}$ = 5453\AA. We have marked some of these wavelengths in Fig A1 showing the $f_{\nu}$ flux (in mags) of Vega between 5400\AA\ and 5600\AA. 

This illustrates the unecessary confusion of these weighted wavelengths. We recommend the retention of only two, the pivot wavelength, $\lambda_{p}$, that is a property of the passband only, and the isophotal wavelength, $\lambda_{iso}$, that takes into account the spectrum measured. To better quantify the derivation of the isophotal wavelength, we recommend that the flux be smoothed to a resolution of 1/3rd of the FWHM of the passband. The pivot wavelength should be used as part of a description of the filter system, while the isophotal wavelength should be used to plot the fluxes as broadband magnitudes against wavelength.  

\begin{figure}
\figurenum{A1}
\plotone{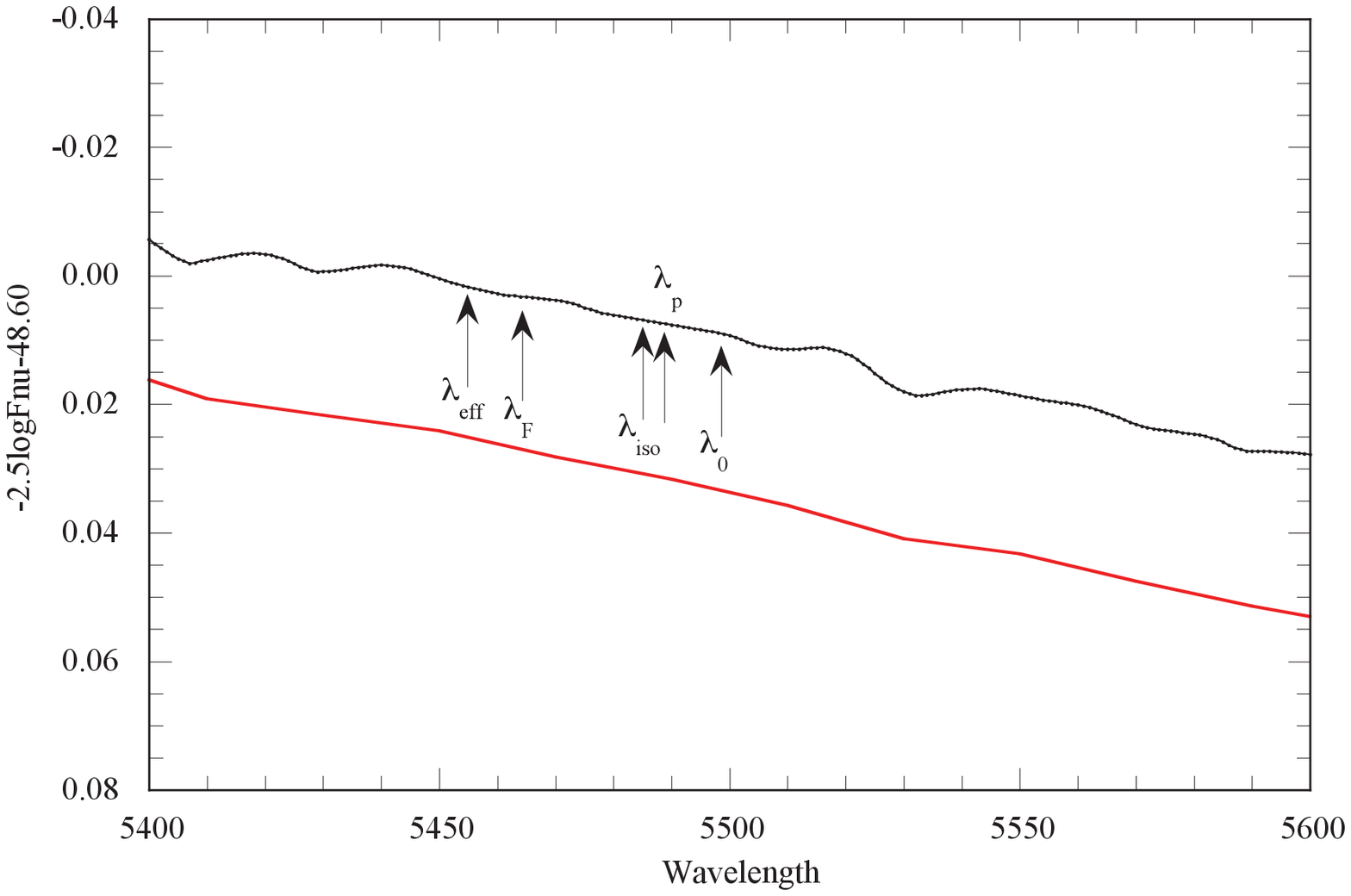}
\caption{STIS005  $f_{\nu}$ spectrum of Vega (black) with some specific wavelengths marked (see text). $\lambda_{F}$ is \citet{Fuku96} $\lambda_{eff}$. The STIS005 spectrum yields $V$ = 0.007 mag. The red line is the Castelli 1994 Vega model flux scaled by $1.2876\times10^{15}$ to produce $V$=0.03 mag. } 
\end{figure}

\subsubsection{Equations involving observed fluxes}
Following \cite{Oke83}, \citet{Fuku96} and \citet{Doi10} defined the broadband AB magnitude as
\begin{equation}
$$ AB mag = $-$2.5 log$\langle f_{\nu}\rangle$ $-$48.60
                     =  $-$2.5 log$\left( \frac{\bigint{f_{\nu}(\nu) S(\nu) d\nu/\nu}} {\bigint{S(\nu) d\nu/\nu}} \right)$ $-$ 48.60
$$
\end{equation}
\citet{Fuku96} imply that this is a photon counting magnitude, however the above equation is the energy integration equation, (see equation A12; and as discussed above, based on the vega\_STIS005 spectrum, the constant should be $-$48.577 to be on the same magnitude scale as the $V$ system for $V_{vega}$ = 0.03). 

Equation A13 showed that photon counting and energy integration magnitudes were equivalent but with different offset constants that are subsumed in the standardisation process. That is, the apparent observed magnitude is usually defined as 
\begin{equation}
$$

$m_{j}$ = $C_{j}$ $-$ 2.5 log $\bigint{n_{p}(\lambda) S(\lambda) d\lambda}$ = $C'_{j}$ $-$ 2.5 log $\bigint{f_{\lambda}(\lambda) S(\lambda) \lambda d\lambda}$
$$
\end{equation}
with the constants $C_{j}$ or $C'_{j}$ found from photometric standards.  Note that the left hand integral describes photon counting while the right hand integral describes energy integration. The regular appearance of the (normalized) product $S(\lambda) \lambda$ in integrals pertaining to photometric magnitudes is often explained simply as wavelength weighting the response function to account for photon counting, however it is primarily a consequence of the modern practice of using the photon response functions rather than the  energy response functions used in the past. 

The most important reason for maintaining the contemporary practice of using photonic response functions is the fact that they are the default response functions in commonly used data reductions packages, such as \textit{synphot} and \textit{pysynphot}. All the passbands published in this paper are photonic passbands $S(\lambda)$. 

\subsubsection{Synphot and pysynphot}
The HST photometry packages \textit{synphot} and \textit{pysynphot} (\url{ http://stsdas.stsci.edu/pysynphot/}; \url{http://stsdas.stsci.edu/stsci\_python\_epydoc/SynphotManual.pdf}) are commonly used for planning HST observations and synthetic photometry. It is useful to relate the definitions, variable names and labels in these packages to those used in this paper. Table A1 is a cross-reference list of terms.

\begin{deluxetable}{llllllllll}
\tabletypesize{\scriptsize}
\tablenum{A1}
\tablecaption{Cross-reference of Photometric Terms }
\tablewidth{0pt}
\tablehead{\colhead{Equation} & \colhead {Name}& \colhead {Description}   & \colhead{synphot } & \colhead {pysynphot}}
\startdata
      & $R$ & pre-detector response function & & &\\
A6 & $S$& photon response function & $P_{\lambda}$ &$P_{\lambda}$  = SpectralElement = bp \\
A9 & $S'$& energy response function &$\lambda P_{\lambda}$\tablenotemark{\dagger} &$\lambda P_{\lambda}$ \\
A14 & $\lambda_{0}$ & mean wavelength & avgwv & bp.avgwave  \\
A16 & $\lambda_{p}$ & pivot wavelength & pivwv & bp.pivot \\
A21 & $\lambda_{eff}$ & effective wavelength &efflam  & Observation.efflam  \\
A25 & $\lambda^{*}_{eff}$& \citet{Schn83} $\lambda_{eff}$ & barlam &  & \\
        & $f_{\lambda}$ &flambda &flam & Observation.effstim('flam')  \\
        &  $f_{\nu}$ &fnu& fnu & Observation.effstim('fnu') \\
A3 & AB$_{\nu}$ & $-$2.5 log $\langle f_{\nu} \rangle$ $-$48.60\tablenotemark{\star} & abmag& Observation.effstim('abmag') \\
A4 & AB$_{\lambda}$ & $-$2.5 log $\langle f_{\lambda} \rangle$ $-$21.10\tablenotemark{\#} & stmag & Observation.effstim('stmag')  \\
\enddata
\tablenotetext{\dagger}{$\lambda P_{\lambda}$ used here is a normalized quantity (made by dividing by the peak wavelength).  }
\tablenotetext{\star}{$-$48.60 \citep{Oke83} used in synphot and pysynphot.}
\tablenotetext{\#}{$-$21.10 \citep{Oke83} used in synphot and pysynphot.}

\end{deluxetable}

\begin{deluxetable}{cccccccccc}
\tabletypesize{\tiny}
\tablenum{1}
\tablecaption{Normalized $UBVRI$ photonic responses}
\tablewidth{0pt}
\tablehead{\colhead{Wave} & \colhead {$U$}&\colhead{Wave} & \colhead {$B$}& \colhead{Wave} & \colhead {$V$}& \colhead{Wave} & \colhead {$R$}&\colhead{Wave} & \colhead {$I$}}
\startdata
3000 & 0.000 & 3600 & 0.000 & 4700 & 0.000 & 5500 & 0.000 & 7000 & 0.000 \\ 
3050 & 0.019 & 3700 & 0.031 & 4800 & 0.033 & 5600 & 0.247 & 7100 & 0.090\\ 
3100 & 0.068 & 3800 & 0.137 & 4900 & 0.176 & 5700 & 0.780 & 7200 & 0.356 \\ 
3150 & 0.167 & 3900 & 0.584 & 5000 & 0.485 & 5800 & 0.942 & 7300 & 0.658 \\ 
3200 & 0.278 & 4000 & 0.947 & 5100 & 0.811 & 5900 & 0.998 & 7400 & 0.865 \\ 
3250 & 0.398 & 4100 & 1.000 & 5200 & 0.986 & 6000 & 1.000 & 7500 & 0.960 \\ 
3300 & 0.522 & 4200 & 1.000 & 5300 & 1.000 & 6100 & 0.974 & 7600 & 1.000 \\ 
3350 & 0.636 & 4300 & 0.957 & 5400 & 0.955 & 6200 & 0.940 & 7700 & 0.998 \\ 
3400 & 0.735 & 4400 & 0.895 & 5500 & 0.865 & 6300 & 0.901 & 7800 & 0.985 \\ 
3450 & 0.813 & 4500 & 0.802 & 5600 & 0.750 & 6400 & 0.859 & 7900 & 0.973 \\ 
3500 & 0.885 & 4600 & 0.682 & 5700 & 0.656 & 6500 & 0.814 & 8000 & 0.970 \\ 
3550 & 0.940 & 4700 & 0.577 & 5800 & 0.545 & 6600 & 0.760 & 8100 & 0.958 \\ 
3600 & 0.980 & 4800 & 0.474 & 5900 & 0.434 & 6700 & 0.713 & 8200 & 0.932 \\ 
3650 & 1.000 & 4900 & 0.369 & 6000 & 0.334 & 6800 & 0.662 & 8300 & 0.904 \\ 
3700 & 1.000 & 5000 & 0.278 & 6100 & 0.249 & 6900 & 0.605 & 8400 & 0.860 \\ 
3750 & 0.974 & 5100 & 0.198 & 6200 & 0.180 & 7000 & 0.551 & 8500 & 0.810 \\ 
3800 & 0.918 & 5200 & 0.125 & 6300 & 0.124 & 7100 & 0.497 & 8600 & 0.734 \\ 
3850 & 0.802 & 5300 & 0.078 & 6400 & 0.075 & 7200 & 0.446 & 8700 & 0.590 \\ 
3900 & 0.590 & 5400 & 0.036 & 6500 & 0.041 & 7300 & 0.399 & 8800 & 0.392 \\ 
3950 & 0.355 & 5500 & 0.008 & 6600 & 0.022 & 7400 & 0.350 & 8900 & 0.203 \\ 
4000 & 0.194 & 5600 & 0.000 & 6700 & 0.014 & 7500 & 0.301 & 9000 & 0.070 \\ 
4050 & 0.107 &           &             & 6800 & 0.011 & 7600 & 0.257 & 9100 & 0.008 \\ 
4100 & 0.046 &           &             & 6900 & 0.008 & 7700 & 0.215 & 9200 & 0.000 \\ 
4150 & 0.003 &           &             & 7000 & 0.006 & 7800 & 0.177 &      &       \\ 
4200 & 0.000 &           &             & 7100 & 0.004 & 7900 & 0.144 &      &       \\ 
     &                  &	  &             & 7200 & 0.002 & 8000 & 0.116 &      &       \\
     &                  &	  &             & 7300 & 0.001 & 8100 & 0.089 &      &       \\
     &                  &	   &            & 7400 & 0.000 & 8200 & 0.066 &      &       \\
     &       &	    &       &      &                                         & 8300 & 0.051 &      &       \\
     &       &	    &       &      &                                         & 8400 & 0.039 &      &       \\
     &       &	    &       &      &                                         & 8500 & 0.030 &      &       \\
     &       &	    &       &      &                                         & 8600 & 0.021 &      &       \\
     &       &	    &       &      &                                         & 8700 & 0.014 &      &       \\
     &       &	    &       &      &                                         & 8800 & 0.008 &      &       \\
     &       &	    &       &      &                                         & 8900 & 0.006 &      &       \\
     &       &	    &       &      &                                         & 9000 & 0.003 &      &       \\
     &       &	    &       &      &                                         & 9100 & 0.000 &      &       \\
\enddata
\end{deluxetable}

\begin{deluxetable}{rrrrrr}
\tabletypesize{\tiny}
\tablenum{2}
\tablecaption{Normalized $H_{p}$ $B_{T}$ $V_{T}$ photonic responses}
\tablewidth{0pt}
\tablehead{\colhead{Wave} & \colhead {$H_{p}$}&\colhead{Wave} & \colhead {$B_{T}$}&\colhead{Wave} & \colhead {$V_{T}$}}
\startdata
3400 & 0.000 & 3500 & 0.000 & 4550 & 0.000 \\
3500 & 0.041 & 3550 & 0.015 & 4600 & 0.023 \\
3600 & 0.072 & 3600 & 0.063 & 4650 & 0.119 \\
3700 & 0.133 & 3650 & 0.132 & 4700 & 0.308 \\
3800 & 0.199 & 3700 & 0.220 & 4750 & 0.540 \\
3900 & 0.263 & 3750 & 0.323 & 4800 & 0.749 \\
4000 & 0.347 & 3800 & 0.439 & 4850 & 0.882 \\
4100 & 0.423 & 3850 & 0.556 & 4900 & 0.951 \\
4200 & 0.508 & 3900 & 0.664 & 4950 & 0.981 \\
4300 & 0.612 & 3950 & 0.751 & 5000 & 0.997 \\
4400 & 0.726 & 4000 & 0.813 & 5050 & 1.000 \\
4500 & 0.813 & 4050 & 0.853 & 5100 & 0.992 \\
4600 & 0.906 & 4100 & 0.880 & 5150 & 0.974 \\
4700 & 0.966 & 4150 & 0.904 & 5200 & 0.946 \\
4800 & 0.992 & 4200 & 0.931 & 5250 & 0.911 \\
4900 & 1.000 & 4250 & 0.960 & 5300 & 0.870 \\
5000 & 0.992 & 4300 & 0.984 & 5350 & 0.827 \\
5100 & 0.978 & 4350 & 1.000 & 5400 & 0.784 \\
5200 & 0.951 & 4400 & 0.969 & 5450 & 0.738 \\
5300 & 0.914 & 4450 & 0.852 & 5500 & 0.692 \\
5400 & 0.880 & 4500 & 0.674 & 5550 & 0.645 \\
5500 & 0.840 & 4550 & 0.479 & 5600 & 0.599 \\
5600 & 0.797 & 4600 & 0.309 & 5650 & 0.553 \\
5700 & 0.755 & 4650 & 0.196 & 5700 & 0.504 \\
5800 & 0.712 & 4700 & 0.131 & 5750 & 0.458 \\
5900 & 0.668 & 4750 & 0.097 & 5800 & 0.412 \\
6000 & 0.626 & 4800 & 0.077 & 5850 & 0.368 \\
6100 & 0.583 & 4850 & 0.056 & 5900 & 0.324 \\
6200 & 0.542 & 4900 & 0.035 & 5950 & 0.282 \\
6300 & 0.503 & 4950 & 0.015 & 6000 & 0.245 \\
6400 & 0.465 & 5000 & 0.003 & 6050 & 0.209 \\
6500 & 0.429 & 5050 & 0.000 & 6100 & 0.178 \\
6600 & 0.393 &  &  & 6150 & 0.152 \\
6700 & 0.359 &  &  & 6200 & 0.129 \\
6800 & 0.326 &  &  & 6250 & 0.108 \\
6900 & 0.293 &  &  & 6300 & 0.092 \\
7000 & 0.260 &  &  & 6350 & 0.078 \\
7100 & 0.230 &  &  & 6400 & 0.066 \\
7200 & 0.202 &  &  & 6450 & 0.055 \\
7300 & 0.176 &  &  & 6500 & 0.044 \\
7400 & 0.152 &  &  & 6550 & 0.036 \\
7500 & 0.130 &  &  & 6600 & 0.027 \\
7600 & 0.112 &  &  & 6650 & 0.017 \\
7700 & 0.095 &  &  & 6700 & 0.008 \\
7800 & 0.081 &  &  & 6750 & 0.000 \\
7900 & 0.068 &  &  &  & \\
8000 & 0.054 &  &  &  & \\
8100 & 0.042 &  &  &  & \\
8200 & 0.032 &  &  &  & \\
8300 & 0.024 &  &  &  & \\
8400 & 0.018 &  &  &  & \\
8500 & 0.014 &  &  &  & \\
8600 & 0.010 &  &  &  & \\
8700 & 0.006 &  &  &  & \\
8800 & 0.002 &  &  &  & \\
8900 & 0.000 &  &  &  & \\
\enddata
\end{deluxetable}

{\textit Facilities:} \facility{HST: STIS}; \facility{INT}; \facility{SSO 2.3m: DBS}
 


\begin{thebibliography}{}
\bibitem[Azusienis \& Straizys(2009)]{Azus69} Azusienis, A., \& Straizys, V., 1969, SvA, 13, 316 
\bibitem[Abazajian et al.(2009)]{Abaz09} Abazajian, K.V. et al. 2009, ApJS, 182, 543
\bibitem[Bessell(1983)]{Bess83} Bessell, M.S., 1983 PASP, 95, 480
\bibitem[Bessell(1986)]{Bess86} Bessell, M.S., 1986, PASP, 98, 1303
\bibitem[Bessell(1990a)]{Bes90a} Bessell, M.S., 1990a, PASP, 102, 1181
\bibitem[Bessell(1990b)]{Bes90b} Bessell, M.S., 1990b, A\&AS, 83, 357
\bibitem[Bessell \& Brett (1988)]{Bess88}Bessell, MS, Brett, JM. 1988. PASP, 100, 1134
\bibitem[Bessell, Castelli \& Plez(1998)]{Bess98} Bessell, M.S., Castelli, F., \& Plez, B., A\&A, 323, 231
\bibitem[Bessell(2000)]{Bess00} Bessell, M.S., 2000, PASP, 112, 961\
\bibitem[Bohlin \& Gilliland(2004)]{Bohl04}Bohlin, R. C., \& Gilliland, R. L. 2004, AJ, 127, 3508
\bibitem[Bohlin(2007)]{Bohl07}Bohlin, R. C. 2007, in The Future of Photometric, Spectrophotometric, and
   Polarimetric Standardization, ASP Conf. Series, Vol. 364, p. 315 ed. C. Sterken
\bibitem[Buser(1986)]{Buse86}Buser, R., 1986, HiA, 7, 799
\bibitem[Buser \& Kurucz(1978)]{Busk78} Buser, R., \& Kurucz, R.L., 1978, A\&A, 70, 555
\bibitem[Cohen et al.(1992)]{Coh92}Cohen, M., Walker, R.G., Barlow, M.J., \& Deacon, J.R., 1992, AJ, 104, 1650
\bibitem[Cousins(1974)]{Cous74} Cousins, A. W. J. 1974, MNRAS, 166, 711 
\bibitem[Cousins(1976)]{Cous76}Cousins, A.W.J., 1976, MmRAS, 81, 25 
\bibitem[Cousins(1984)]{Cous84}Cousins, A.W.J., 1984, SAAOCirc, 8, 69
\bibitem[Cousins \& Menzies(1993)]{Cous93}Cousins, A.W.J., Menzies, J.W., 1993, in Precision Photometry. Proceedings of a conference held to honour A.W.J. Cousins in his 90th year, held Observatory, Cape Town, South Africa, 2-3 February 1993. Edited by D. Kilkenny, E. Lastovica and J.W. Menzies. Cape Town: South African Astronomical Observatory (SAAO), p.240 
\bibitem[Doi et al.(2010)]{Doi10}Doi, M., Tanaka, M., Fukugita, M., Gunn, J.E., Yasuda, N., Ivezic, Z., Brinkmann, J., de Haars, E., Kleinman, S.J., Krzesinski, J., French Leger, R., 2010, AJ,139, 1628 
\bibitem[Fukugita et al.(1996)]{Fuku96}Fukugita, M., Ichikawa, T., Gunn, J.E., Doi, M., Shimasaku, K., Schneider, D.P., 1996, AJ, 111, 1748
\bibitem[Gray(2007)]{Gray07}Gray, R. O. 2007, in The Future of Photometric, Spectrophotometric, and
   Polarimetric Standardization, ASP Conf. Series, Vol. 364, p. 305 ed. C. Sterken
\bibitem[Grenon(2001)]{Gren01} Grenon, M., 2001, private communication
\bibitem[Gronbech \& Olsen(1976)]{Gron76}Gronbech, B., Olsen, E.H., 1976, A\&AS, 34, 1
\bibitem[Gustafsson et al.(2008)] {Gust08} Gustafsson, B., Edvardsson, B., Eriksson, K., Jorgensen, U.G., Nordlund, A., Plez, B., 2008, A\&A, 486, 951
\bibitem[Hayes(1985)]{Hay85}Hayes, D.S., 1985, In IAU Symposium 111: Calibration of fundamental stellar quantities, ed. D.S. Hayes, L.E. Pasinetti and A.G.Davis Philip, (Reidel: Dordrecht), p. 225
\bibitem[Heap \& Lindler(2007)]{Heap07} Heap, S.R., \& Lindler, D. 2007, IAUS, 241, 95 
\bibitem[Hog et al.(2000)]{Hog00} Hog E., Fabricius C., Makarov V.V., Urban S., Corbin T., Wycoff G., Bastian U., Schwekendiek P., Wicenec A., ``The Tycho2 catalogue", 2000, A\&A, 355, L27
\bibitem[Johnson et al.(1966)]{John66}Johnson, H.L., Iriarte, B., Mitchell, R.I., Wisniewskj, W.Z., 1966, CmLPL, 4, 99
\bibitem[Kaiser et al.(2010)]{Kais10}Kaiser, N,  Burgett, W., Chambers, K., Denneau, L., Heasley, J., Jedicke, R., Magnier, E., Morgan, J., Onaka, P. \& Tonry, J., 2010, SPIE 7733, 77330E1
\bibitem[Keller et al.(2007)]{Kell07}Keller, S. et al., 2007, PASA, 24, 1
\bibitem[Kerzendorf(2011)]{Kerz11} Kerzendorf, W., 2011, private communication
\bibitem[King(1952)]{King52}King, I., 1952, ApJ, 115, 580
\bibitem[Kilkenny et al.(1998)]{Kilk98}Kilkenny, D., van Wyk, F., Roberts, G., Marang, F., \& Cooper, D., 1998, MNRAS, 294, 93
\bibitem[Koen et al.(2002)]{Koen02}Koen, C., Kilkenny, D., van Wyk, F., Cooper, D., \& Marang, F., 2002, MNRAS, 334, 20
\bibitem[Koen et al.(2010)]{Koen10}Koen, C., Kilkenny, D., van Wyk, F., \& Marang, F., 2010, MNRAS, 403, 1949
\bibitem[Koornneef et al.(1986)]{Koor86}Koornneef, J., Bohlin, R., Buser, R, Horne, K., Turnshek, D., 1986, HiA, 7, 833
\bibitem[Landolt(1983)]{Land83} Landolt, A.U., 1983, AJ, 88, 439
\bibitem[Landolt \& Uomoto(2007)]{Land07} Landolt, A.U., Uomoto, A.K., 2007, AJ, 133, 768
\bibitem[Landolt(2009)]{Land09} Landolt, A.U., 2009, AJ, 137, 4186
\bibitem[Maiz Appellaniz(2006)]{Maiz06} Maiz Apellaniz, J., 2006, AJ, 131, 1184
\bibitem[Megessier(1995)]{Mege95} Megessier, C., 1995, A\&A, 296, 771
\bibitem[Menzies et al.(1989)]{Menz89} Menzies, J. W., Cousins, A. W. J, Banfeld, R. M., \& Laing, J. D., 1989, SAAO Circ., 13, 1
\bibitem[Menzies(1990)]{Menz90} Menzies, J. W., 1990, private communication
\bibitem[Menzies(1993)]{Menz93}Menzies, J.W., 1993, in Precision Photometry. Proceedings of a conference held to honour A.W.J. Cousins in his 90th year, held Observatory, Cape Town, South Africa, 2-3 February 1993. Edited by D. Kilkenny, E. Lastovica and J.W. Menzies. Cape Town: South African Astronomical Observatory (SAAO), p.35
\bibitem[Mermilliod (2006)]{Merm06} Mermilliod, J.C., 2006 Vizier II/168; Mermiliiod (1991) Universite de Lausanne.
\bibitem[Munari et al.(2005)]{Muna05}Munari, U., Sordo, R., Castelli, F., Zwitter, T., 2005, A\&A, 442, 1127
\bibitem[Nicolet(1978)]{Nico78} Nicolet, B., 1978, A\&AS, 34, 1
\bibitem[Nicolet(1996)]{Nico96}Nicolet, B. 1996. BaltA, 5, 417
\bibitem[Olsen(1983)]{Olse83}Olsen, E.H., 1983, A\&AS, 54, 55
\bibitem[Oke(1965)]{Oke65} Oke, J.B., 1965, ARAA, 3,  23
\bibitem[Oke \& Schild(1970)]{Oke70} Oke, J.B. \& Schild, R., E. 1970, ApJ, 161, 1015
\bibitem[Oke \& Gunn(1983)]{Oke83} Oke, J.B. \& Gunn, J.E., 1983, ApJ, 266, 713
\bibitem[Pel(1990)]{Pell90} Pel, J-W., 1990, private communication 
\bibitem[Pel \& Lub(2007)]{Pell07} Pel, J-W.,\& Lub, J., 2007, ASPConf, 364, 63 
\bibitem[Perryman et al.(1997)]{Perr97} Perryman, M.A.C., Lindegren, L., Kovalevsky, J., Hog, E., Bastian, U., Bernacca, P.L., Creze, M., Donati, F., Grenon, M., Grewing, M., van Leeuwen, F., van der Marel, H., Mignard, F., Murray, C.A., Le Poole, R.S., Schrijver, H., Turon, C., Arenou, F., Froeschle, M., Petersen, C.S., ``The Hipparcos Catalogue", 1997,A\&A,323,L49
\bibitem[Rufener \& Nicolet(1988)]{Rufe88} Rufener, F. \& Nicolet, B., 1988, A\&A, 206, 357
\bibitem[Sanchez-Blazquez et al.(2006)]{Sanc06}Sanchez-Blazquez, P., Peletier, R. F., Jimenez-Vicente, J., Cardiel, N., Cenarro, A. J., Falcon-Barroso, J., Gorgas, J., Selam, S., Vazdekis, A., 2006,MNRAS,371,703
\bibitem[Schneider et al.(1983)]{Schn83}Schneider, D.P., Gunn, J.E., \& Hoessel, J.G., 1983, ApJ, 264, 337
\bibitem[Soffer \& Lynch(1999)]{Soff99} Soffer, B.H., \& Lynch, D.K., 1999, AmJPhys, 67, 946
\bibitem[Strai\v{z}ys(1996)]{Strai96}Strai\v{z}ys, V. 1996. BaltA, 5, 459	
\bibitem[Straizys \& Sviderskiene(1972)]{Stra72} Straizys, V. \& Sviderskiene, Z., 1972, Astron. Obs. Bull. Vilnius, 35, 1
\bibitem[Tokunaga \& Vacca(2005)]{Toku05}Tokunaga, A.T., \& Vacca, W.D., 2005, PASP, 117, 421
\bibitem[van Leeuwen et al.(1997a)]{vLee97a} van Leeuwen, F., Lindegren, L., Mignard, F., 1997, ÔThe Hipparcos and Tycho CataloguesÕ Vol. 3, ESA SP-1200, 461.
\bibitem[van Leeuwen et al.(1997b)]{vLee97b} van Leeuwen, F., Evans, D. W., Grenon, M., Grossmann, V., Mignard, F., Perryman, M. A. C., 1997, A\&A, 323, L61
\end{thebibliography}
\end{document}